\documentclass[epj]{svjour}
\RequirePackage{graphicx}
\RequirePackage{mathptmx}      
\RequirePackage{flushend}
\RequirePackage[numbers,sort&compress]{natbib}
\RequirePackage[colorlinks,citecolor=blue,urlcolor=blue,linkcolor=blue]{hyperref}
\RequirePackage{amsmath,amssymb,latexsym}
\RequirePackage{siunitx}

\newcommand{\vare}{\varepsilon}

\newcommand{\pt}{\partial}

\newcommand{\msun}{M_\odot}

\begin{document}

\title{Bayesian modeling of the nuclear equation of state for neutron star tidal deformabilities and GW170817}

\author{Y.\ Lim\inst{1}
	\and
	J.\ W.\ Holt\inst{1,2} 
}


\institute{Cyclotron Institute, Texas A\&M University, College Station, TX 77843, USA\label{1}
	\and
	Department of Physics and Astronomy, Texas A\&M University, College Station, TX 77843, USA\label{2}
}

\date{Received: date / Accepted: date}

\abstract{
We present predictions for neutron star tidal deformabilities obtained from a Bayesian analysis of the nuclear equation of state, assuming a minimal model at high-density that neglects the possibility of phase transitions. The Bayesian posterior probability distribution is constructed from priors obtained from microscopic many-body theory based on realistic two- and three-body nuclear forces, while the likelihood functions incorporate empirical information about the equation of state from nuclear experiments. The neutron star crust equation of state is constructed from the liquid drop model, and the core-crust transition density is found by comparing the energy per baryon in inhomogeneous matter and uniform nuclear matter. From the cold $\beta$-equilibrated neutron star equation of state, we then compute neutron star tidal deformabilities as well as the mass-radius relationship. Finally, we investigate correlations between the neutron star tidal deformability and properties of finite nuclei.
}

\maketitle


\section{Introduction}

The recent observation of gravitational wave event GW170817 has led to numerous inferred constraints on the nuclear equation of state (EOS) and bulk neutron star properties, such as radii, tidal deformabilities, and moments of inertia \cite{Fattoyev18,annala18,most18,krastev18,lim18a,tews2018gw,tsang18,landry18,abbott2018b}. Observations of the associated electromagnetic counterpart, and in particular the energetics of the resulting kilonova, suggest that the neutron star merger remnant was a long-lived hypermassive neutron star \cite{bauswein17,margalit17,shibata17,radice18,rezzolla18,ruiz18}, which itself places strong constraints on the dense matter equation of state and in particular the maximum mass for nonrotating neutron stars. Anticipated multi-messenger observations of future neutron star merger events have the potential to further refine such constraints, and in addition, the simultaneous measurement of neutron star radii and masses expected from the NICER mission \cite{dimitrios14} will complement these ongoing gravitational wave searches.

During a neutron star merger event, the late-inspiral gravitational wave signal contains information on the tidal deformation induced from the gravitational field of one neutron star on the other. For a given neutron star mass, the tidal deformability is strongly correlated with the matter radius. Observation of the post-merger peak frequency, which characterizes the primary oscillation mode of the hypermassive neutron star remnant, can also provide constraints on neutron star radii \cite{bauswein12}. Current inferred upper limits on neutron star radii and tidal deformabilities from GW170817 rule out stiff equations of state that result in large radii ($R_{1.4} \gtrsim 13.6$\,km) and large tidal deformabilities ($\Lambda_{1.4} \gtrsim 700$). 

In the past, neutron star tidal deformabilities have been studied within the framework of Skyrme Hartree-Fock theory \cite{Hinderer2010,Postnikov2010,malik18,kim2018}, relativistic mean field theory \cite{kumar17,Fattoyev18,malik18}, chiral effective field theory (EFT) coupled with extrapolations to high density \cite{most18,lim18a,tews2018gw}, and polytropic equations of state \cite{raithel18,zhao18}. Since these different theoretical models have quite large uncertainties, especially with respect to the high-density equation of state, most have been shown to be consistent with the recent re-analysis \cite{abbott2018b,de2018a} of the neutron star tidal deformabilities determined from GW170817. This implies that current observational data is not yet capable of favoring one model approach over another, as long as the derived EOS obtained can produce relatively soft equations of state with small neutron star tidal deformabilities. What we can get from the various equations of state consistent with GW170817 are nuclear matter properties at and above nuclear saturation density ($n_0=0.16$\,fm$^{-3}$), especially the pressure at around two to three times nuclear saturation density. This is natural since the central density of a $1.4\,\msun$ neutron star has a distribution that peaks around three times nuclear saturation density for soft equations of state, as we show in more detail below.

In the simplest approximation one can regard a neutron star as a giant nucleus containing $\sim 10^{57}$ nucleons. In that case, one can derive from simple liquid drop model arguments the typical radius of a neutron star:
\begin{equation}
R \simeq r_0 A^{1/3} \simeq 12\,\mathrm{km},
\end{equation}
where $r_0=1.2\,\mathrm{fm}$. Neutron stars, however, are considerably more complex, with phase changes occurring with increasing density. In particular the outer crust consists of a lattice of ionized nuclei in a gas of ultra-relativistic electrons. The inner crust consists of neutron-rich nuclei in a lattice together with unbound electrons and superfluid neutrons. Finally, the core consists of uniform matter containing minimally protons, neutrons, and electrons, with possible novel states of matter in the inner core, including hyperons, deconfined quark matter, and meson condensates \cite{baym73,au74,glen82,glen91,thorsson94,glen98,bunta04,weber05,alford05,brown07,brown08,weissenborn11,weissenborn12a,weissenborn12b,lim14,lim15h,lim17h}. Since the average central density of neutron stars with $M>1.4\,\msun$ is beyond three times nuclear saturation density, the inter-nucleon spacing $d=\mathrm{0.78}$\,fm at $n=3n_0$ is smaller than the proton charge radius, $0.84\sim 0.88$\,fm \cite{carlson15}. Thus, there are strong motivations for hypothesized hadron-quark phase transitions in neutron star inner cores. Even before such high densities, hyperons with chemical potentials lower than that of nucleons are expected to appear, though there are still significant uncertainties associated with hyperon-nucleon-nucleon three-body forces \cite{Lonardoni:2013rm,Petschauer:2016pbn,Haidenbauer:2016vfq} that may delay the onset of hyperonic matter in neutron stars.

Besides the matter composition in neutron stars, the nuclear equation of state is crucial for calculating the macroscopic structure, mass-radius relation, tidal deformability, moment of inertia and other bulk properties of neutron stars \cite{lattimer01,lattimer05,lattimer12nd,yagi13,lim18b}. Thus, constructing the nuclear EOS has been a primary challenge in dense matter research. In the low-density region, the neutron star EOS is constrained entirely by the properties of finite nuclei. Upon reaching the inner crust, nuclear properties such as neutron skin thicknesses and the EOS of dilute neutron matter (available from microscopic many-body calculations) help constrain neutron star structure. The properties of dense nuclear matter beyond about twice saturation density are strongly model dependent and there is no theoretical framework that gives controlled uncertainty estimates in this regime. Intermediate-energy heavy-ion collisions \cite{danielewicz02,tsang18} allow for the experimental investigation of matter at these densities, but resulting constraints on the equation of state are model dependent with large uncertainties. Thus, astrophysical observations of neutron stars are expected to provide possibly the strongest constraints on the properties of matter at supra-saturation density. Such observational data include the maximum mass for nonrotating neutron stars ($M_\mathrm{max} > 1.97M_\odot$) \cite{Demorest:2010bx,Antoniadis:2013pzd}, the tidal deformability from gravitational wave signals \cite{abbott17a}, the mass-radius relation from X-ray burst data \cite{SLB2010,SLB2016,ozel16,Bogdanov16,suleimanov16,nattil17}, the potential measurement of neutron star moments of inertia \cite{lyne04,lattimer05,landry18,lim18b}, and simultaneous mass and radius measurements from NICER \cite{dimitrios14}.
Recently, the measurement of $M = 2.17^{+0.11}_{-0.10}\,\msun$ at the $1\sigma$ credibility level for PSR J0774+6620 by Cromartie {\it et al}.\ \cite{Cromartie19} may result in an increase of the lower bound on the maximum neutron star mass, thereby excluding 
some nuclear force models.

In the present work, we outline a method for including the latest constraints on the dense matter equation of state from microscopic modeling of nuclear and neutron matter together with empirical information about the equation of state from laboratory measurements of finite nuclei. The method is based on Bayesian analysis where prior probability distributions for the EOS model parameters are obtained from chiral effective field theory calculations at low to moderate densities and likelihood functions that incorporate empirical properties of medium-mass and heavy nuclei. From the resulting posterior probability distributions for the model parameters, we then compute neutron star tidal deformabilities, radii, and correlations among bulk neutron star properties and equation of state parameters, such as the symmetry energy and its slope parameter. The paper is organized as follows. In Section\,\ref{sec:model} we explain the nuclear models we employ for dense matter as well as our Bayesian methodology to generate the equation of state with clear statistical interpretation. In Section \ref{sec:tidal} we briefly explain how to compute neutron star tidal deformabilities from the Tolman-Oppenheimer-Volkoff (TOV) equation. In Section\,\ref{sec:res} we present numerical results from our parametrized equations of state and study the correlation between neutron star quantities and nuclear matter properties. We summarize our results in Section\,\ref{sec:sum}.


\section{Nuclear Modeling}
\label{sec:model}

\subsection{Equation of state}

The equation of state for cold beta-equilibrated nuclear matter is required over a very large range of conditions (density and particle composition) not accessible to laboratory experiments on Earth. Thus it is necessary to extrapolate the EOS to highly isospin-asymmetric systems and to densities beyond nuclear matter saturation density. A common framework is to employ polytropic equations of state \cite{hebeler10prl,hebeler13,raithel16,raithel17}, which have the freedom to change the adiabatic index and account for the possibility of phase transitions at specified densities. In addition, Skyrme Hartee-Fock or relativistic mean field (RMF) models \cite{stone03,Steiner2005325,fattoyev10,lim14,lim2017a,lim17}, whose parameters are fitted to the properties of finite nuclei close to saturation density, are generally extrapolated to much higher densities in order to study the mass-radius relationship for neutron stars.

It has been shown \cite{krueger13,ermal16} that many mean field model calculations are not consistent with the low-density equation of state for pure neutron matter constrained by chiral effective field theory \cite{hebeler10,drischler14,drischler16,holt17prc,Sammarruca18}. Chiral effective field theory is formally a well converged expansion for strongly interacting systems when the characteristic momentum scale is well below the chiral symmetry breaking scale $\Lambda_\chi$ $\simeq 1$\,GeV, but in practice nuclear potentials impose a smaller cutoff scale $\Lambda \simeq 400-600$\,MeV to regulate the high-momentum components of the interaction. The maximum density at which the theory may provide controlled theoretical uncertainty estimates is therefore between $1-2n_0$. While quantum Monte Carlo (QMC) calculations with realistic two- and three-body nuclear potentials have been frequently employed in the past to study the properties of pure neutron matter \cite{gandolfi2012,gezerlis13,roggero14,wlazlowski14,tews16}, the symmetric nuclear matter equation of state is more challenging \cite{gandolfi07} due to the low-density spinodal instability and the larger number of nucleons needed to fill doubly-closed-shell box eigenstates. Thus, many-body perturbation theory has been the most widely used method to compute the equation of state for asymmetric nuclear matter. These calculations and the associated theoretical uncertainties can then be used to constrain nuclear energy density functionals (EDF) for which the energy density $\mathcal{E}$ of homogeneous matter is given as a function of the baryon number density $n$ and proton fraction $x=n_p/(n_n+n_p)$.

Chiral effective field theory suggests that a natural expansion parameter for the energy density is $k_F$, which is proportional to $n^{1/3}$ at $T=0$\,MeV. We therefore parametrize a set of energy density functionals according to the following form
\begin{equation}
\label{eq:edf}
\begin{aligned}
\mathcal{E}(n,x) & =  \frac{1}{2m}\tau_n + \frac{1}{2m}\tau_p \\
& + [1-(1-2x)^2] f_s(n) + (1-2x)^2 f_n(n) \,,
\end{aligned}
\end{equation} 
where $\tau_p$ ($\tau_n$) denotes the kinetic energy density of protons (neutrons), and $f_s$ ($f_n$) corresponds to the potential energy density of symmetric nuclear matter (neutron matter):
\begin{equation}
\label{eq:fns}
f_s(n) = \sum_{i=0}^{3} A_i\, n^{(2+i/3)} \,, 
\quad
f_n(n) = \sum_{i=0}^{3} B_i\,n^{(2+i/3)}\,.
\end{equation}
For convenience we rewrite the above expressions formally as expansions about a reference Fermi momentum $k_F^r$:
\begin{equation}
\label{eq:exp}
\begin{aligned}
\frac{\mathcal{E}}{n}(k_F,x=0.5) &= 2^{2/3}\frac{3}{5}\frac{k_F^2}{2m} + \frac{k_F^3}{9\pi^2} \sum_{i=0}^{3} \frac{a_i}{i!}\, \beta^i \\
\frac{\mathcal{E}}{n}(k_F,x=0) & = \frac{3}{5}\frac{k_F^2}{2m} + \frac{k_F^3}{9\pi^2} \sum_{i=0}^{3} \frac{b_i}{i!}\, \beta^i\,,
\end{aligned}
\end{equation}
where $\beta = (k_F - k_F^r) / k_F^r$ and in both expansions we define $k_F = ( 3 \pi ^ 2 n ) ^ {1/3}$. In the next section we will perform a global fit of the equations of state from chiral effective field theory to the form given in Eq.\ \eqref{eq:exp}, and therefore the choice of reference Fermi momentum is unimportant. In the present case we take $k_F^r = 1.68$\,fm$^{-1}$, corresponding to the neutron matter Fermi momentum at saturation density.

In previous work \cite{holt18} the authors have shown that the above ansatz can well describe the density dependence of the nuclear isospin-asymmetry energy $S_2(n)$ computed in chiral effective field theory from low-densities up to twice saturation density. The normal Taylor series expansion of the isospin-asymmetry energy around saturation density
\begin{equation}
\label{symtay}
\begin{aligned}
S_2(n) & = J + L \left(\frac{n-n_0}{3n_0} \right) 
           + \frac{1}{2}K_{\mathrm{sym}}\left(\frac{n-n_0}{3n_0} \right)^2 \\
       & \phantom{=J}\,  + \frac{1}{6}Q_{\mathrm{sym}}\left(\frac{n-n_0}{3n_0} \right)^3
            + \cdots        
\end{aligned}
\end{equation}
generally does not describe well the low-density behavior since there is nothing to enforce $S_2(n) \rightarrow 0$ as $n \rightarrow 0$. Here we do not account for the effects of clustering, which would lead to finite values of $S_2$ as the density approaches zero \cite{natowitz10}. For the EDFs employed in this work, the effective masses of neutrons and protons are embedded in the functional forms since the kinetic momentum term $\tau$ is proportional to $n^{5/3}$ at $T=0$\,MeV in uniform nuclear matter.

\begin{figure}[t]
\centering
\includegraphics[scale=0.33]{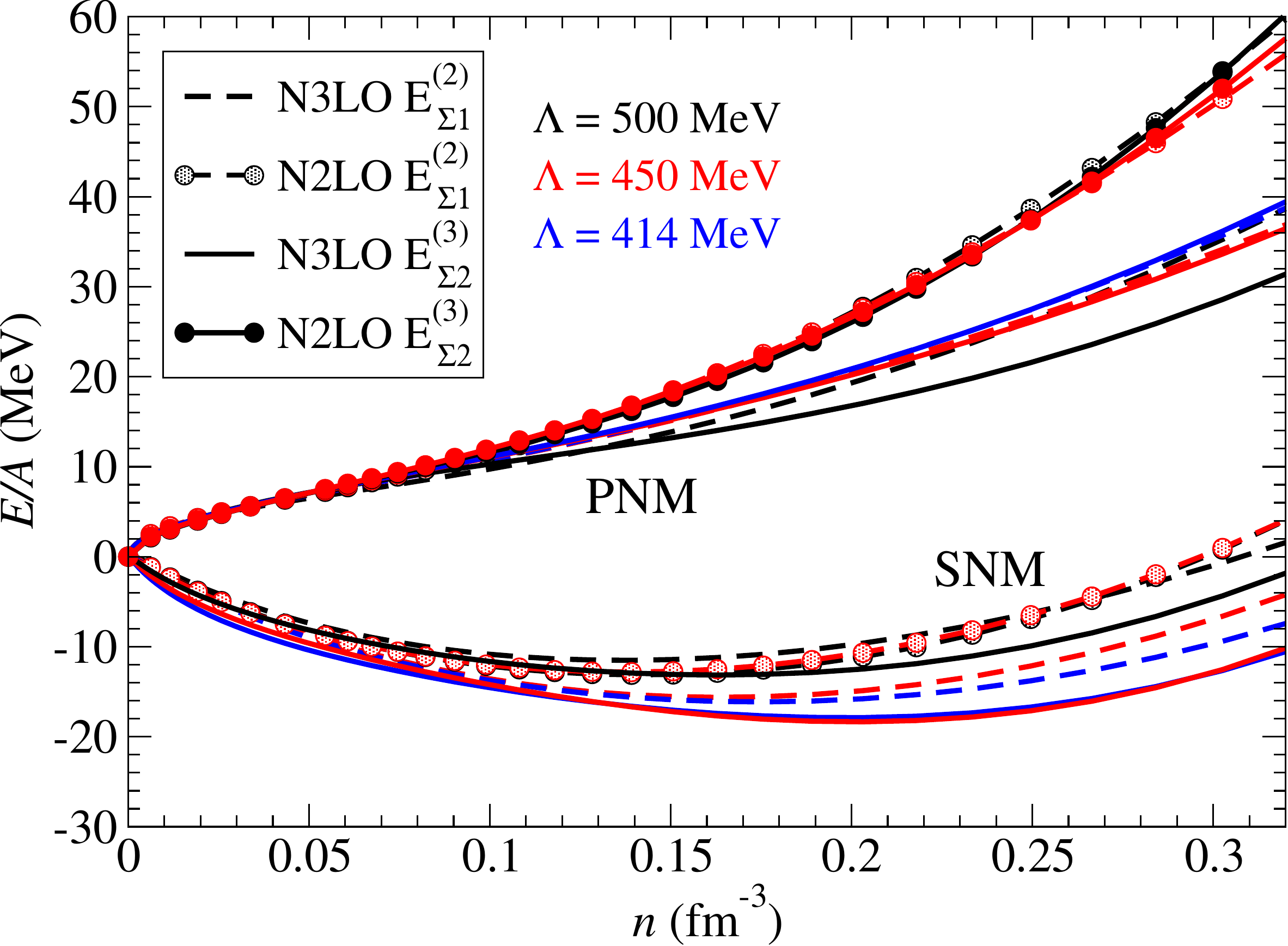}
\caption{Zero-temperature equation of state for pure neutron matter (PNM) and symmetric nuclear matter (SNM) calculated from chiral two- and three-nucleon forces in many-body perturbation theory at second order (dashed lines) and third order (solid lines) with N2LO (circles) and N3LO nucleon-nucleon potentials. In all cases the N2LO chiral three-body force is included.}
\label{fig:eftfit}
\end{figure}

In Fig.\ \ref{fig:eftfit} we show the energy per baryon for pure neutron matter and symmetric nuclear matter from chiral two- and three-body forces \cite{coraggio14,sammarruca15}. We choose three values for the momentum cutoff scale: $\Lambda = 414$\,MeV (blue), $450$\,MeV (red), and $500$\,MeV (black). The minimum value $\Lambda = 414$ MeV is the relative momentum corresponding to the lab energy $E=350$\,MeV for which nucleon-nucleon (NN) elastic scattering phase shift data is typically incorporated into fits of high-precision nucleon-nucleon interactions. In addition to the cutoff scale, we also vary the order in the chiral expansion, where next-to-next-to-leading order (N2LO) chiral NN potentials are denoted with circles, while N3LO NN potentials have no symbols. In all cases, we include the N2LO chiral three-body force whose low-energy constants are fitted to the binding energies of $A=3$ nuclei together with the lifetime of $^3$H. To estimate the theoretical uncertainty from many-body perturbation theory, we calculate the energy per particle at second order with intermediate-state energies in the Hartree-Fock approximation (dashed lines, $E^{(2)}_{\Sigma 1}$) and at third order with self-consistent intermediate-state energies at second order (solid lines, $E^{(3)}_{\Sigma 2}$). These results from chiral effective field theory are fitted up to $n=2n_0$ to the form in Eqs.\ \eqref{eq:edf} and \eqref{eq:fns} with correlation coefficients $R\simeq 0.9999$. This minimal energy density functional is also useful to fit theoretical calculations from quantum Monte Carlo as well as Skyrme or Gongy effective interactions, unless phase transitions to exotic matter are involved. In the following analysis, these predictions from chiral effective field theory will determine the Bayesian prior probability distributions for the $a_i$ and $b_i$ parameters of Eq.\ (\ref{eq:exp}).

It is often assumed that the energy density around symmetric nuclear matter ($x=1/2$) can be expanded using a Maclaurin series:
\begin{equation}
\frac{\mathcal{E}(n,x)}{n} 
= \sum_{i=0}^{\infty}S_{2i}(n)(1-2x)^{2i}\,.
\label{isomac}
\end{equation}
In fact, it has been shown \cite{kaiser15,wellenhofer16} that the series does not generically converge due to the existence of logarithmic terms that appear beyond the mean field approximation, leading to the more general expansion
\begin{equation}
\begin{aligned}
\frac{\mathcal{E}(n,x)}{n} 
= & S_0(n) + S_2(n)(1-2x)^2 \\
& + \sum_{i=2}^{\infty}(S_{2i} + L_{2i} \ln \vert 1-2x \vert )(1-2x)^{2i}\,.
\end{aligned}
\label{isodep}
\end{equation}
Regardless of whether Eq.\ \eqref{isomac} or \eqref{isodep} is considered, it has been shown \cite{lagaris81,Bombaci91,wellenhofer16,drischler16b} that the contributions beyond $S_2$ are small when computed microscopically and can be neglected. We therefore assume a quadratic dependence on the isospin asymmetry in the energy density functional in Eq.\ \eqref{eq:edf}. 

We use four parameters to describe the symmetric nuclear matter and pure neutron matter energy densities. In the case of symmetric nuclear matter, the set of $\{a_i\}$ parameters can be constrained from empirical properties of medium-mass and heavy nuclei, and in particular the values of the nuclear saturation energy $B$, saturation density $n_0$, incompressibility $K$, and skewness $Q$ defined at $n=n_0$ and $x=1/2$:
\begin{equation}
\begin{aligned}
& B = - \frac{\mathcal{E}(n_0)}{n_0}\,, \quad
\left . p = n_0^2\frac{\partial (\mathcal{E}/n)}{\pt n}\right |_{n=n_0}=0\,, \\
& \left . K = 9 n_0^2\frac{\partial^2 (\mathcal{E}/n)}{\pt n^2}\right |_{n=n_0}\,,\quad
\left . Q = 27n_0^3\frac{\partial^3 (\mathcal{E}/n)}{\pt n^3}\right |_{n=n_0}\,,
\end{aligned}
\end{equation}
where $p$ is the pressure of symmetric nuclear matter at saturation density. In the case of pure neutron matter, we can consider empirical constraints on the parameters $J$, $L$, $K_{\rm sym}$, and $Q_{\rm sym}$ defined in Eq.\ \eqref{symtay} in order to obtain the set of $\{b_i\}$ in Eq.\ \eqref{eq:exp}. However, compared to the parameters of the symmetric nuclear matter equation of state around saturation density, the pure neutron matter empirical parameters have much larger uncertainties. We therefore employ experimental constraints on the symmetry energy $J$ together with correlations among $J$, $L$, $K_{\rm sym}$, and $Q_{\rm sym}$ found in recent works \cite{lattimer13,tews17,holt18,margueron18eob}.


\subsection{Bayesian statistics}
In Bayesian analysis, the posterior probability distribution for a vector of model parameters $\mathbf{a}$ is obtained as the product of the likelihood function for a set of data given the parameter distribution $\mathbf{a}$ and the prior distribution function for $\mathbf{a}$ that incorporates previous knowledge \cite{Sivia06}. More precisely,
\begin{equation}
P(\mathbf{a}\vert \mathrm{data}) \sim P(\mathrm{data}\vert \mathbf{a})P(\mathbf{a}),
\label{bayes}
\end{equation}
where $P(\mathbf{a}\vert \mathrm{data})$ is the posterior distribution, $P(\mathrm{data}\vert \mathbf{a})$ is the likelihood function, and $P(\mathbf{a})$ is the prior distribution. In the present case $\mathbf{a}=(a_i, b_i)$. For the present purposes, the explicit normalization of the posterior probability distribution does not need to be specified.

\begin{table}[t]
	\caption{Covariance matrix for the $\{a_i\}$ parameters associated with the symmetric nuclear matter equation of state from chiral effective field theory.}
	\begin{center}
		\begin{tabular}{c|cccc}
			\hline
			& $a_0$  &  $a_1$ & $a_2$  & $a_3$ \\
			\hline
			$a_0$ & $0.05$\,fm$^4$  &    $0.06$\,fm$^4$  &    $0.09$\,fm$^4$  &  $-4.01$\,fm$^4$  \\
			\hline
			$a_1$ & $0.06$\,fm$^4$  &    $0.20$\,fm$^4$  &    $0.31$\,fm$^4$  &   $-9.11$\,fm$^4$  \\
			\hline
			$a_2$ & $0.09$\,fm$^4$  &    $0.31$\,fm$^4$  &    $1.87$\,fm$^4$  &  $-16.35$\,fm$^4$  \\
			\hline
			$a_3$ & $-4.01$\,fm$^4$  &   $-9.11$\,fm$^4$  &  $-16.35$\,fm$^4$  &  $696.53$\,fm$^4$ \\
			\hline
		\end{tabular}
	\end{center}
	\label{covsnm}
\end{table}

\begin{table}[t]
	\caption{Covariance matrix for the $\{b_i\}$ parameters associated with the pure neutron matter equation of state from chiral effective field theory.}
	\begin{center}
		\begin{tabular}{c|cccc}
			\hline
			& $b_0$  &  $b_1$ & $b_2$  & $b_3$ \\
			\hline
			$b_0$ & $0.04$\,fm$^4$  &  $0.19$\,fm$^4$  &  $0.40$\,fm$^4$  &  $0.21$\,fm$^4$ \\
			\hline
			$b_1$ & $0.19$\,fm$^4$  &  $0.94$\,fm$^4$  &  $2.07$\,fm$^4$  &   $-0.20$\,fm$^4$ \\
			\hline
			$b_2$ & $0.40$\,fm$^4$  &  $2.07$\,fm$^4$  &    $9.26$\,fm$^4$ &  $30.23$\,fm$^4$ \\
			\hline
			$b_3$ & $0.21$\,fm$^4$  & $-0.20$\,fm$^4$  & $30.23$\,fm$^4$ & $227.42$\,fm$^4$ \\
			\hline
		\end{tabular}
	\end{center}
	\label{covpnm}
\end{table}

Since the free parameters in chiral nuclear forces are fitted to only the properties of $A=2,3$ nuclei, results for the nuclear matter equation of state are theoretical predictions. We have therefore suggested in previous work \cite{lim18a} that the prior probability distributions for $\{a_i\}$ and $\{b_i\}$ be obtained by fitting the equations of state from chiral effective field theory to the form in Eqs.\ \eqref{eq:edf} and \eqref{eq:fns}. We fit each equation of state individually and then compute the mean vectors and covariance matrices for the $\{a_i\}$ and $\{b_i\}$ parameter sets independently. We obtain for the means $\{\bar a_i\} = \{-3.48$\,fm$^2, 6.15$\,fm$^2, -1.51$\,fm$^2, 39.58$\,fm$^2$\} and $\{\bar b_i\} = \{-1.70$\,fm$^2, 3.87$\,fm$^2, 4.61$\,fm$^2, 16.85$\,fm$^2$\}. Results for the covariance matrices are shown in Tables \ref{covsnm} and \ref{covpnm}.

\begin{figure}
\centering
\includegraphics[scale=0.5]{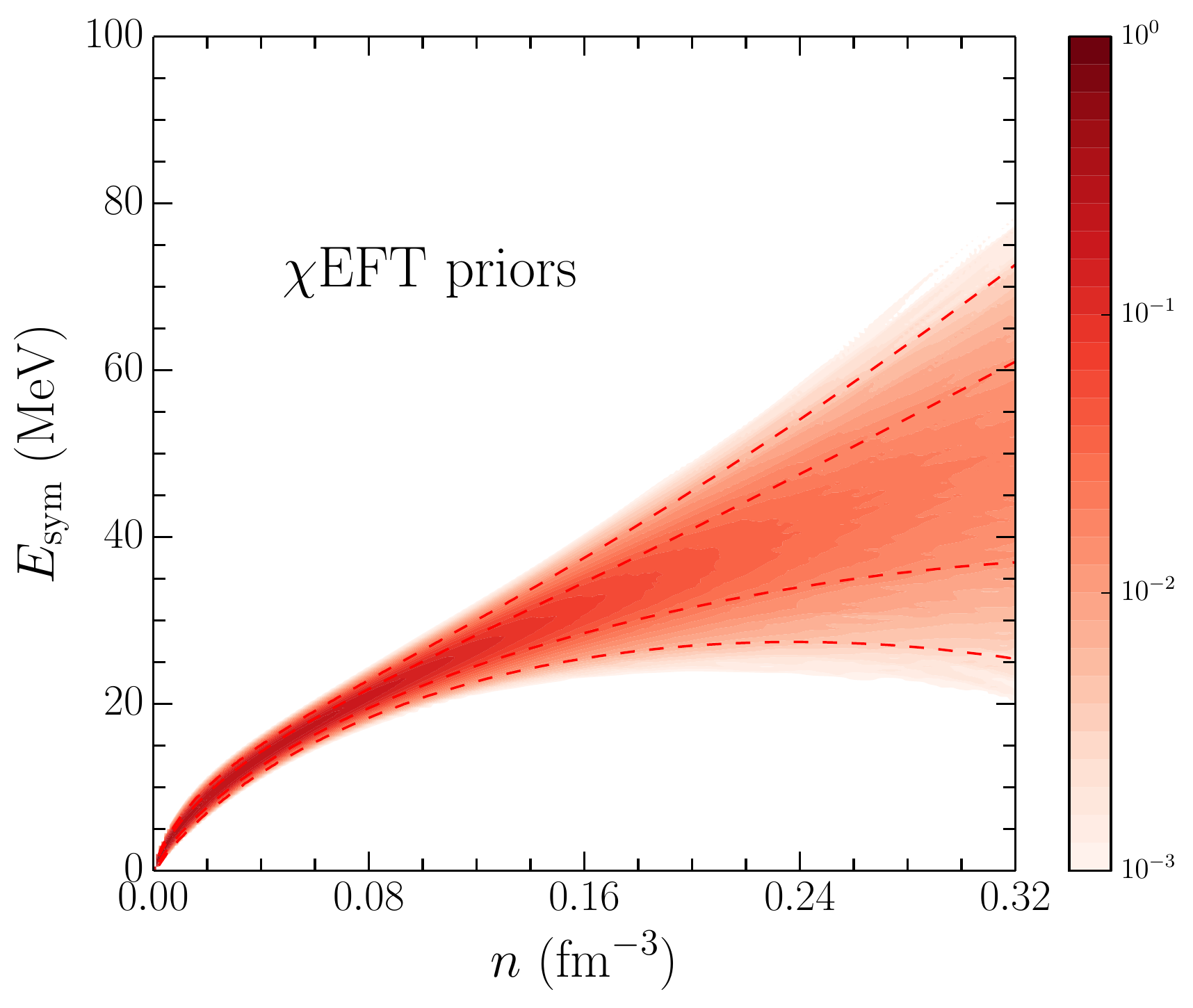}
\caption{Density-dependent symmetry energy $E_{\rm sym}$ from the prior probability distributions for $\{a_i\}$ and $\{b_i\}$. The dashed lines indicate the $1\sigma$ and $2\sigma$ uncertainty bands.}
\label{fig:esym_chi}
\end{figure}

In Fig.\ \ref{fig:esym_chi} we show the resulting probability distribution for the nuclear symmetry energy $E_{\rm sym}$ defined as the difference between the energy per nucleon of pure neutron matter and symmetric nuclear matter at a given density. We sample from Gaussian prior distributions for the $\{a_i\}$ and $\{b_i\}$ parameters from chiral effective field theory. In Fig.\ \ref{fig:esym_chi} the dashed lines correspond to the $1\sigma$ and $2\sigma$ probability contours. We observe that the theoretical uncertainties on the nuclear symmetry energy grow rapidly with the nuclear density, reaching $\Delta E_{\rm sym} \simeq 50$\,MeV at $n=2n_0$.

\begin{figure}[t]
	\centering
	\begin{tabular}{cc}
		\includegraphics[scale=0.21]{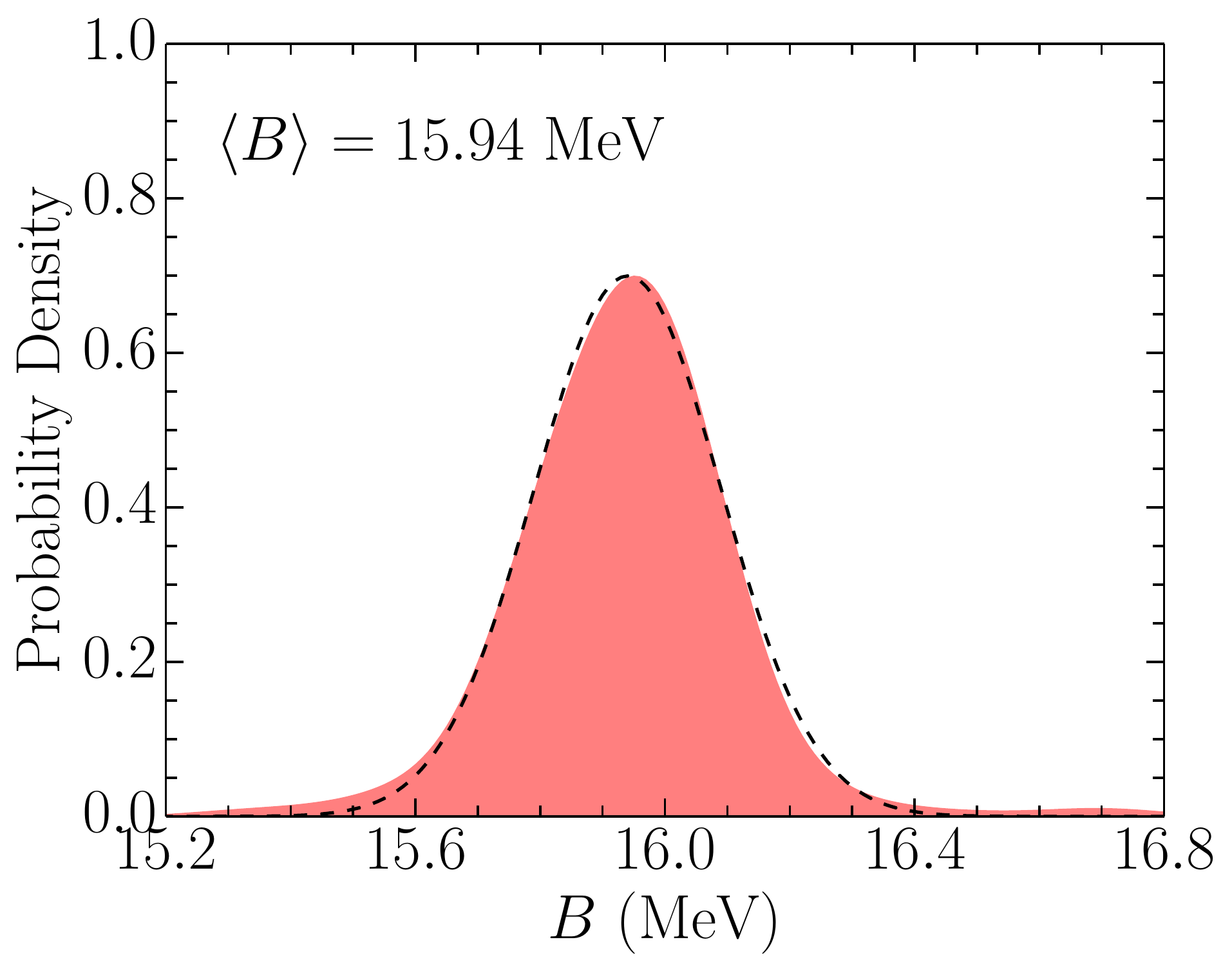} &
		\includegraphics[scale=0.21]{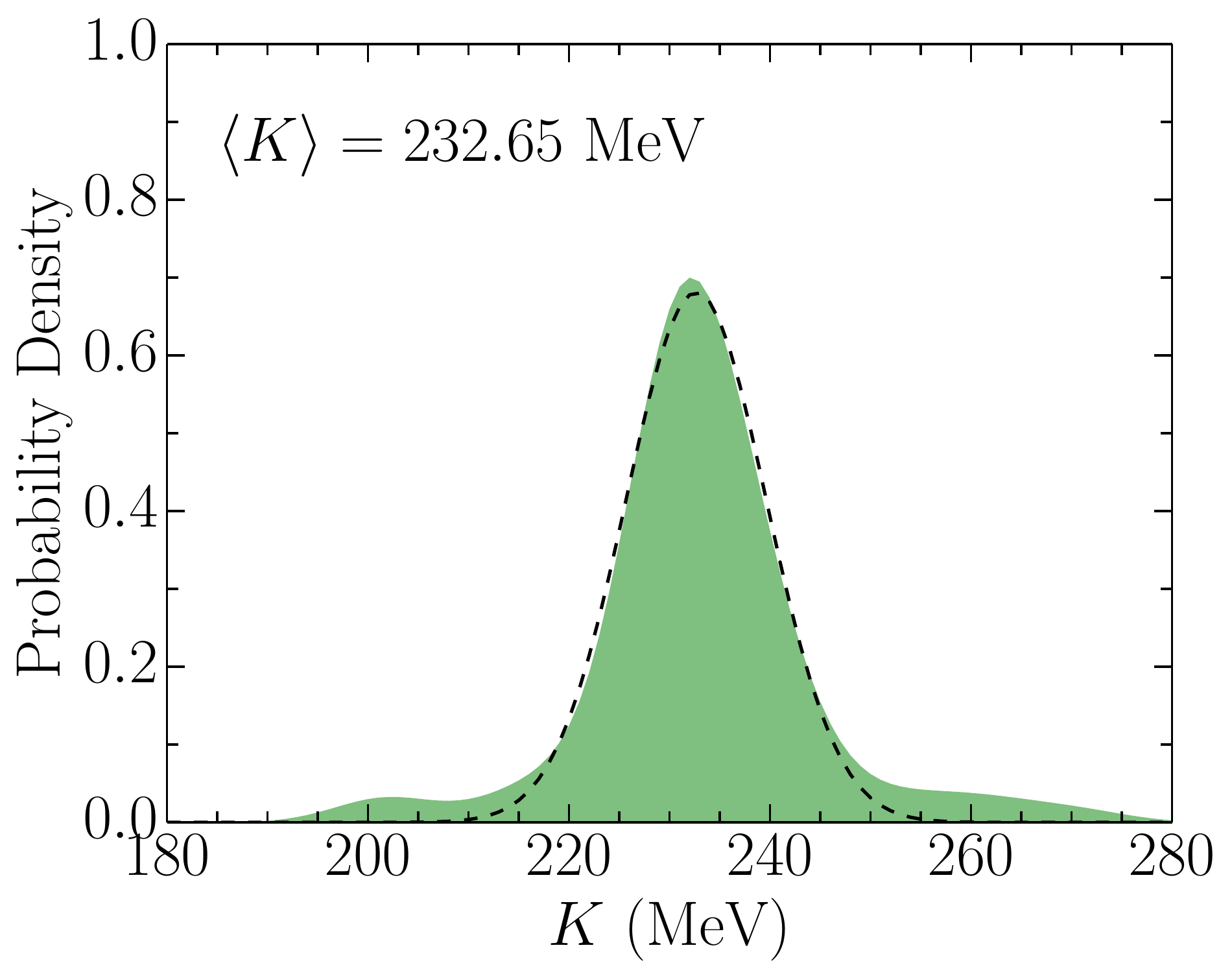} \\
		\includegraphics[scale=0.21]{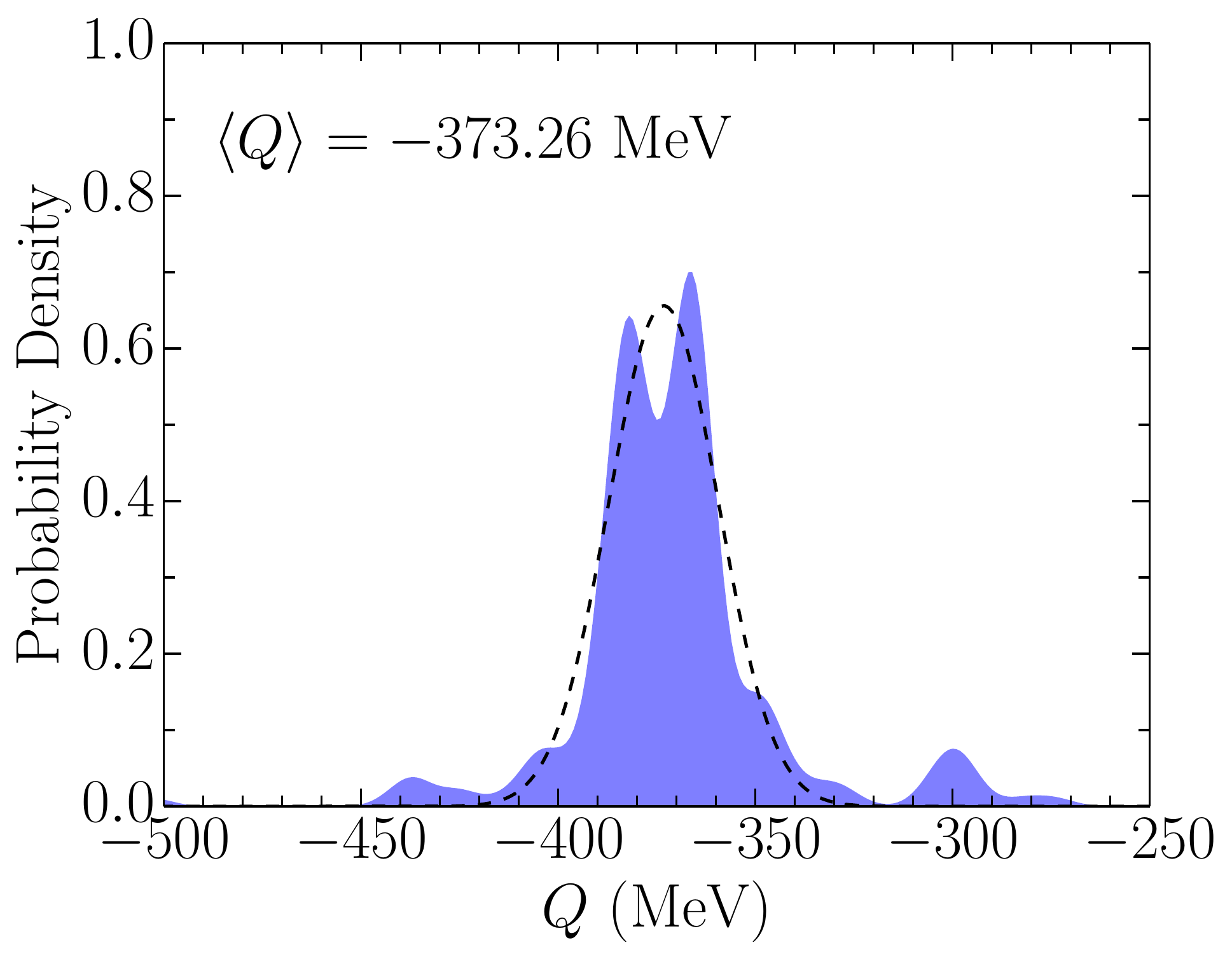} &
		\includegraphics[scale=0.21]{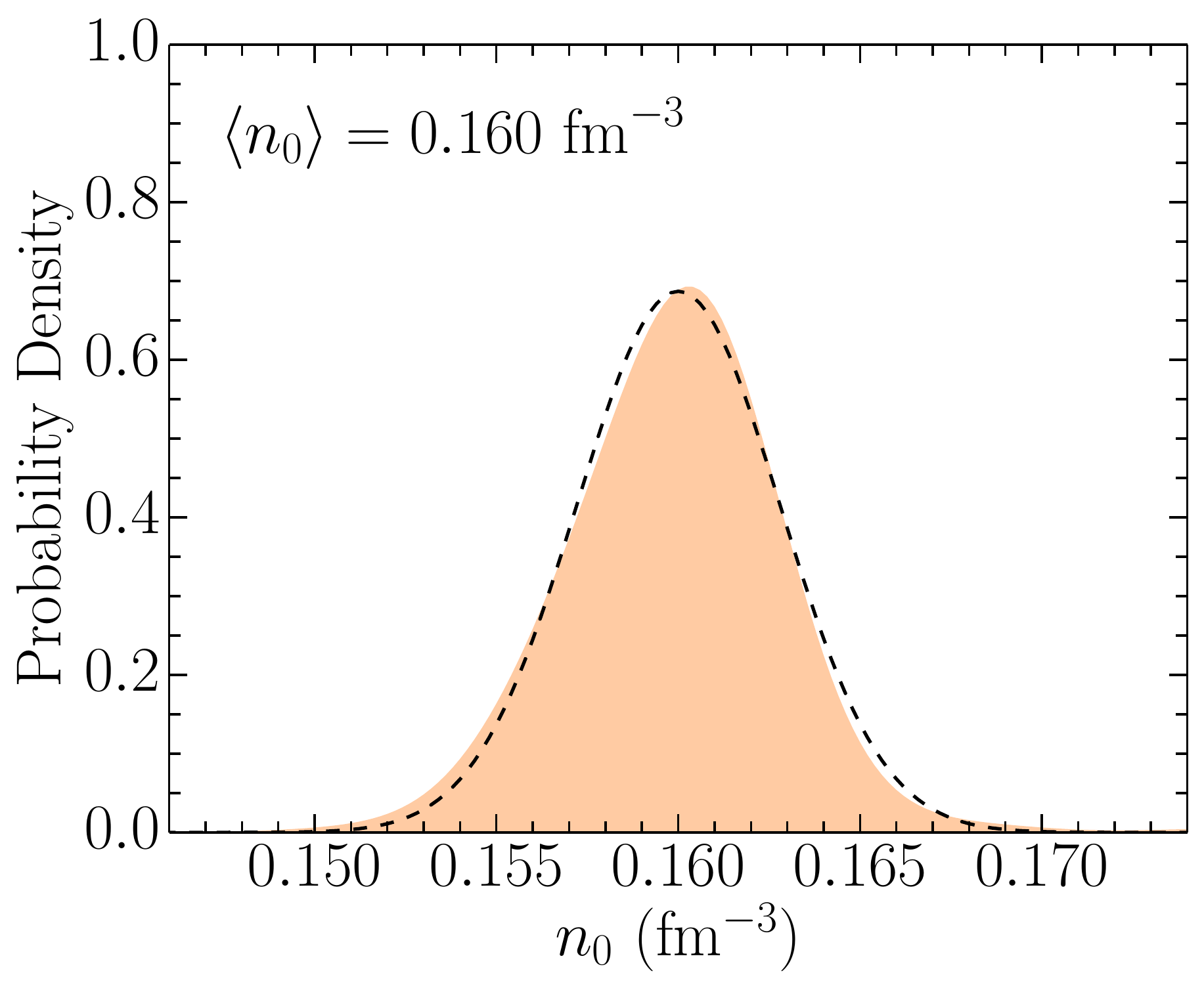}
	\end{tabular}
	\caption{Distributions of the symmetric nuclear matter empirical parameters $B, n_0, K$, and $Q$ obtained from 205 Skyrme force models. Also shown are the best-fit Gaussian distribution functions for each quantity and their mean values.}
	\label{fig:skyrmeall}
\end{figure}

We next discuss how to include experimental data, such as the binding energies and charge radii of medium-mass and heavy nuclei, into likelihood functions involving the $\{a_i\}$ and $\{b_i\}$ parameters. For the $\{a_i\}$ parameters entering the symmetric nuclear matter equation of state, we employ a large set of 205 Skyrme effective interactions benchmarked to the properties of nuclear matter in Ref.\ \cite{dutra12}. In particular, we consider four symmetric nuclear matter properties $B, n_0, K, Q$ and compute the distributions of these quantities obtained from the 205 Skyrme interactions. In Fig.\ \ref{fig:skyrmeall} we show these statistical distributions together with Gaussian fits. From the explicit relationship between these empirical nuclear matter properties and the coefficients $a_i$ in our Fermi momentum expansion in Eq.\ \eqref{eq:exp}, we then derive a joint likelihood function involving the $\{a_i\}$. Using just the individual distributions of $B$, $n_0$, $K$, $Q$ to find the $a_i$ parameters for symmetric nuclear matter would neglect the correlations among those quantities. Therefore, it is important to take into account the full covariance matrices involving the $a_i$.

\begin{figure}[t]
	\centering
	\begin{tabular}{cc}
		\includegraphics[scale=0.21]{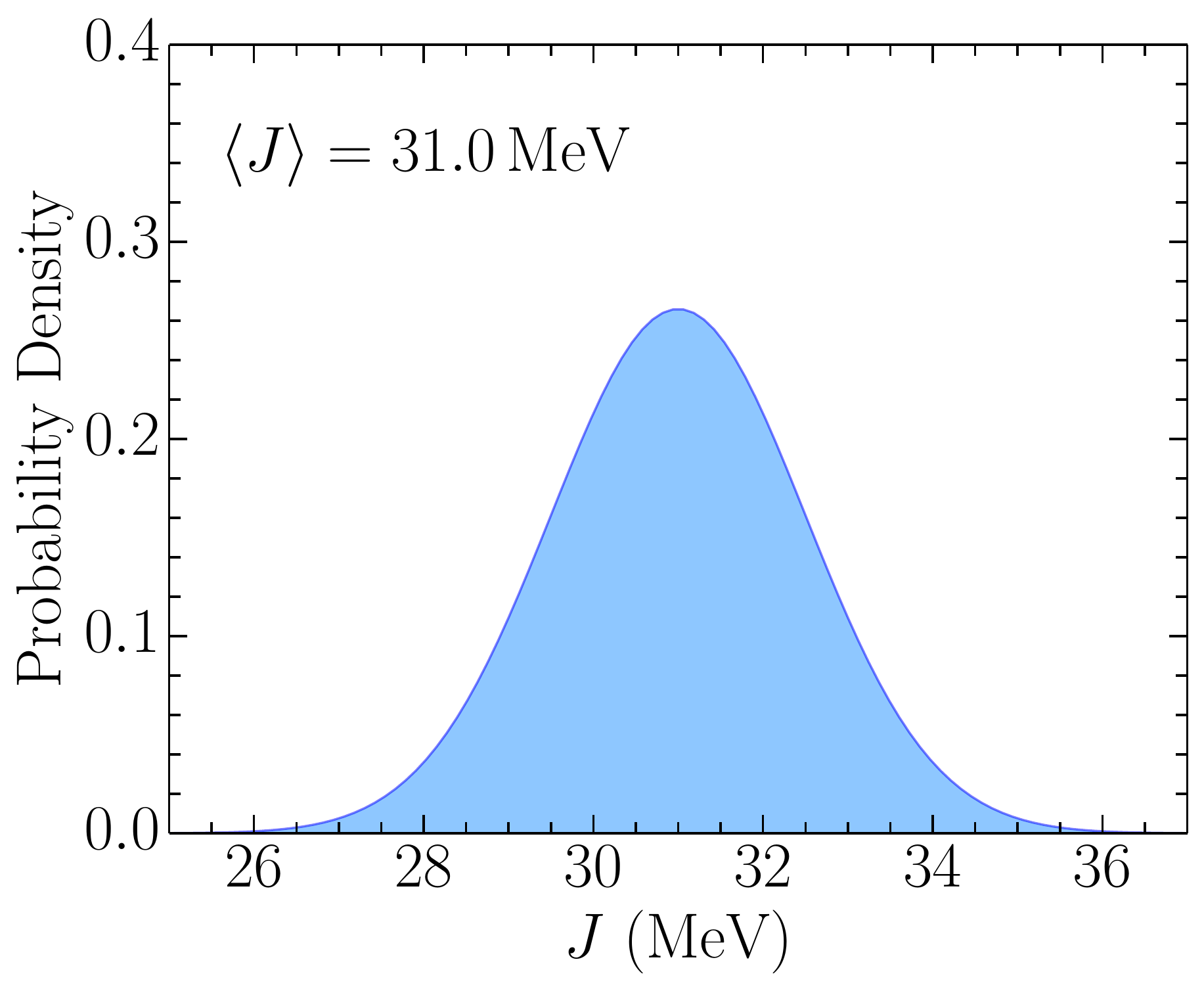} &
		\includegraphics[scale=0.21]{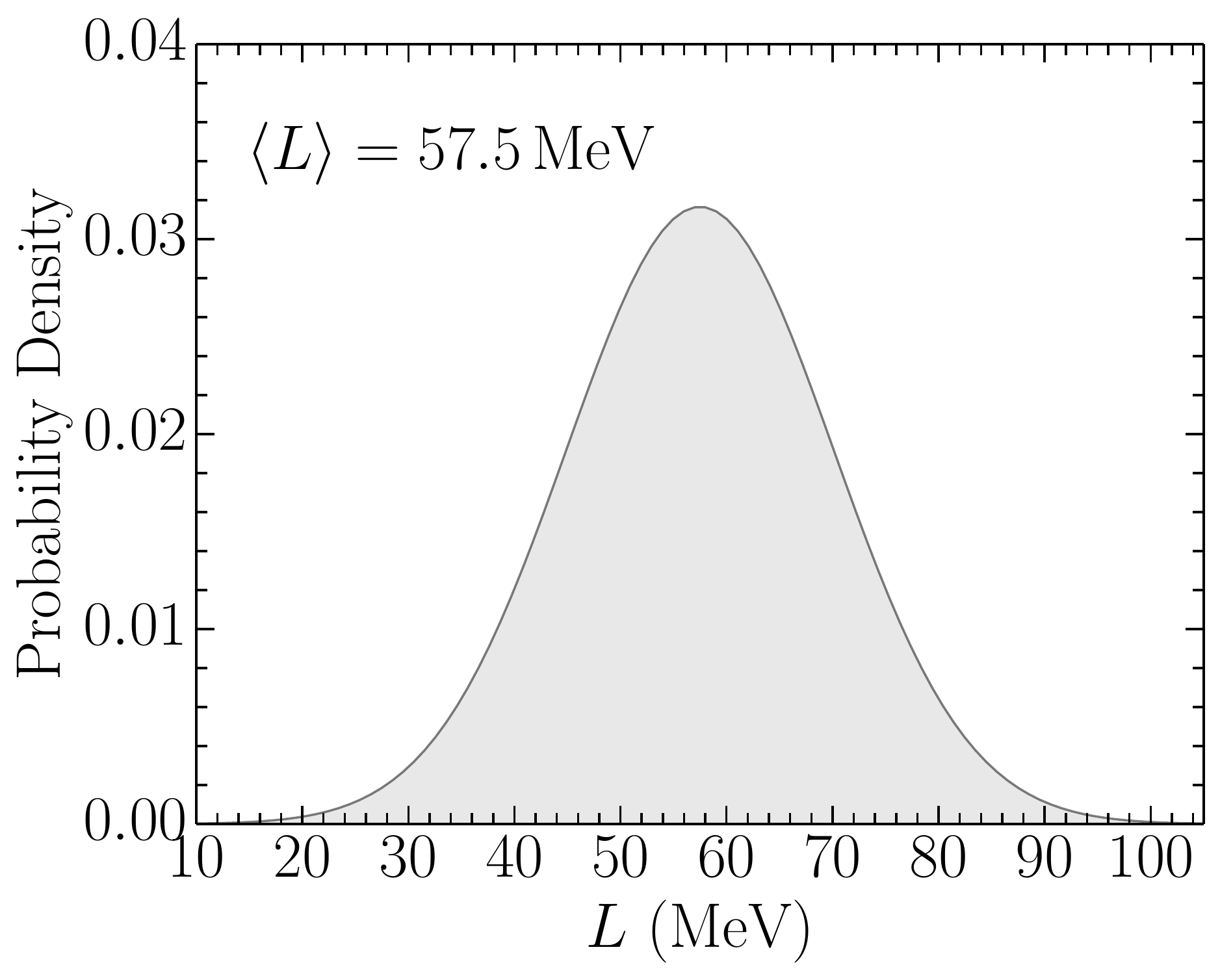} \\
		\includegraphics[scale=0.21]{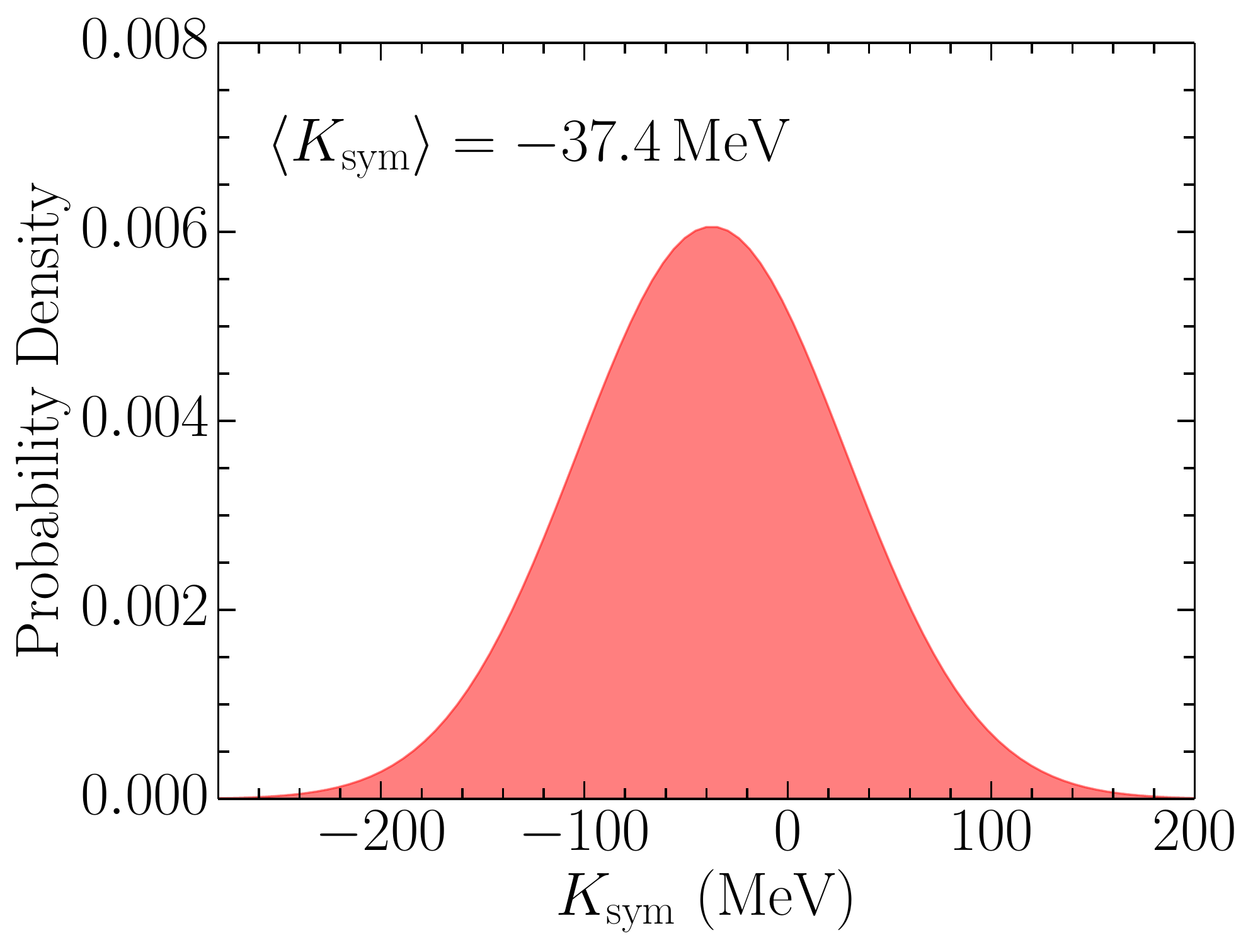} &
		\includegraphics[scale=0.21]{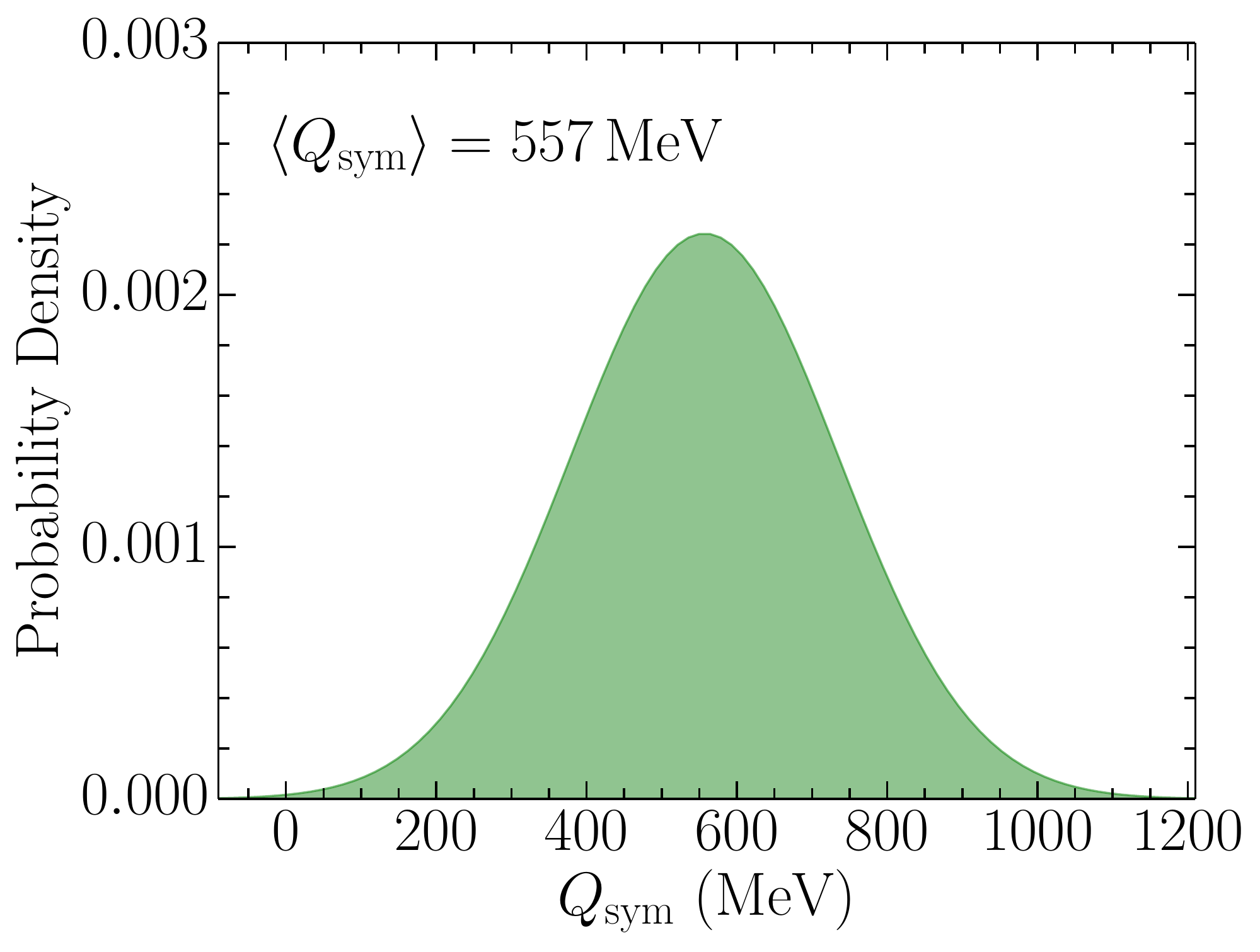}
	\end{tabular}
	\caption{Probability densities for the symmetry energy parameters $J, L, K_{\rm sym}, Q_{\rm sym}$ associated with the Bayesian likelihood function involving the $b_i$ parameters in Eq.\ \eqref{eq:exp}. The distributions are obtained from the empirical bound on the symmetry energy at saturation density, $J = 31 \pm 1.5$\,MeV, together with correlations between $J$ and $L, K_{\rm sym}, Q_{\rm sym}$.}
	\label{fig:bi}
\end{figure}

For the pure neutron matter equation of state, we employ empirical data on the symmetry energy at saturation density $J$, together with correlations among $J$, $L$, $K_{\rm sym}$, and $Q_{\rm sym}$. In particular, in Ref.\ \cite{holt18} it was shown that $L$ and $J$ are linearly correlated, as well as $K_{\rm sym}$ and $J$. For a linear correlation of the form $Z = pX + Y$, where $X$ and $Y$ are uncorrelated, we have
\begin{equation}
{\rm cov}(Z,X) = p\, {\rm var}(X).
\end{equation}
This relationship allows us to extract from the correlation bands among $J$, $L$, and $K_{\rm sym}$ in Ref.\ \cite{holt18} the mean and covariance matrix elements for the $b_i$ parameters. Since correlations between $Q_{\rm sym}$ and the other symmetry energy parameters were not considered in Ref.\ \cite{holt18}, we simply extract from their parametrization a broad distribution for the $b_3$ neutron matter coefficient. In Fig.\ \ref{fig:bi} we show the resulting distributions for the symmetry energy empirical parameters $J, L, K_{\rm sym}, Q_{\rm sym}$.

To generate an ensemble of equations of state, we sample from the $a_i$ and $b_i$ posterior probability distributions obtained as a product of the prior probability distribution (determined from microscopic chiral EFT calculations) and likelihood function (determined from empirical information about medium-mass and heavy nuclei). We have assumed Gaussian prior probability distributions for the $a_i$ and $b_i$ parameters obtained from microscopic modeling of the dense matter equation of state. Since the likelihood functions in Figs.\ \ref{fig:skyrmeall} and \ref{fig:bi} are also approximately Gaussian, the values of $a_i$ and $b_i$ can be generated randomly from the posterior probability distribution according to the average of each variable and weighted by the covariance matrix for each variable. 

Since all equation of state constraints come from the region $n \leq 2n_0$, our modeling at high densities is limited and does not explore the widest range of theoretical scenarios, such as phase transitions, hyperons, or meson condensates. The description of neutron star properties described below should therefore be interpreted as a minimal model. In particular, the behavior of the nuclear equation of state for dense nuclear matter beyond twice saturation density is assumed to follow the energy density functional in Eq.\ \eqref{eq:exp}. Thus, the polytropic slope does not evolve in the density range $n>2n_0$. In the future we plan to allow for the possibility of phase transitions and higher powers of the Fermi momentum in Eq.\ (\ref{eq:exp}). In the present work we generate 300,000 samples each for symmetric nuclear matter and pure neutron matter. For arbitrary proton fractions we interpolate between the symmetric nuclear matter and pure neutron matter equations of state, keeping only the term in Eq.\ \eqref{isodep} proportional to the square of the isospin asymmetry $\delta_{np} = (n_n-n_p)/(n_n+n_p) = 1 - 2x$.

\begin{figure}[t]
	\centering
	\includegraphics[scale=0.45]{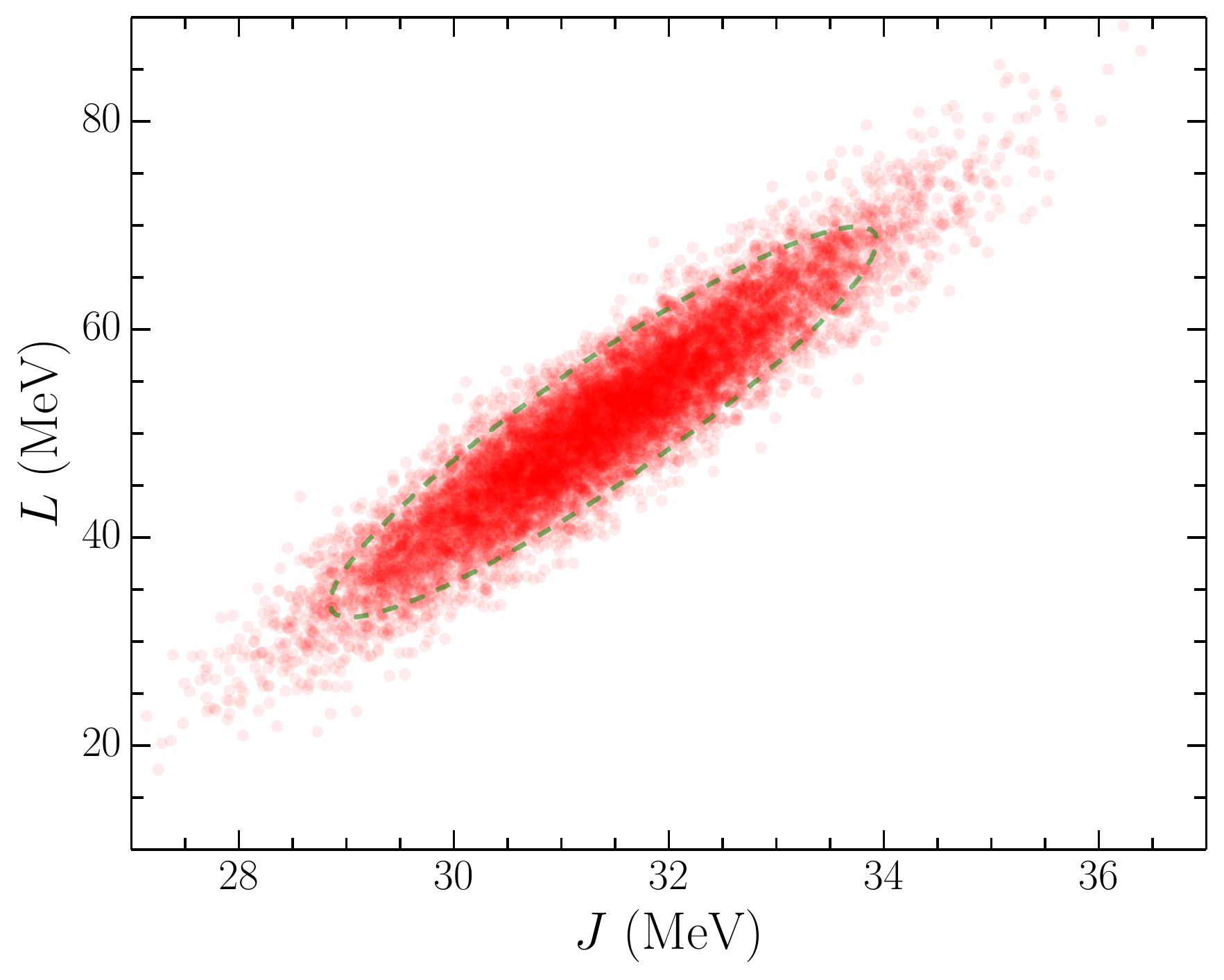}
	\caption{Distribution for $J$ and $L$ of the energy density functionals generated from the posterior probability distributions in this work. The dashed line denotes the $2\sigma$ correlation ellipse.}
	\label{fig:corrsvl}
\end{figure}

\begin{figure}
\centering
\includegraphics[scale=0.44]{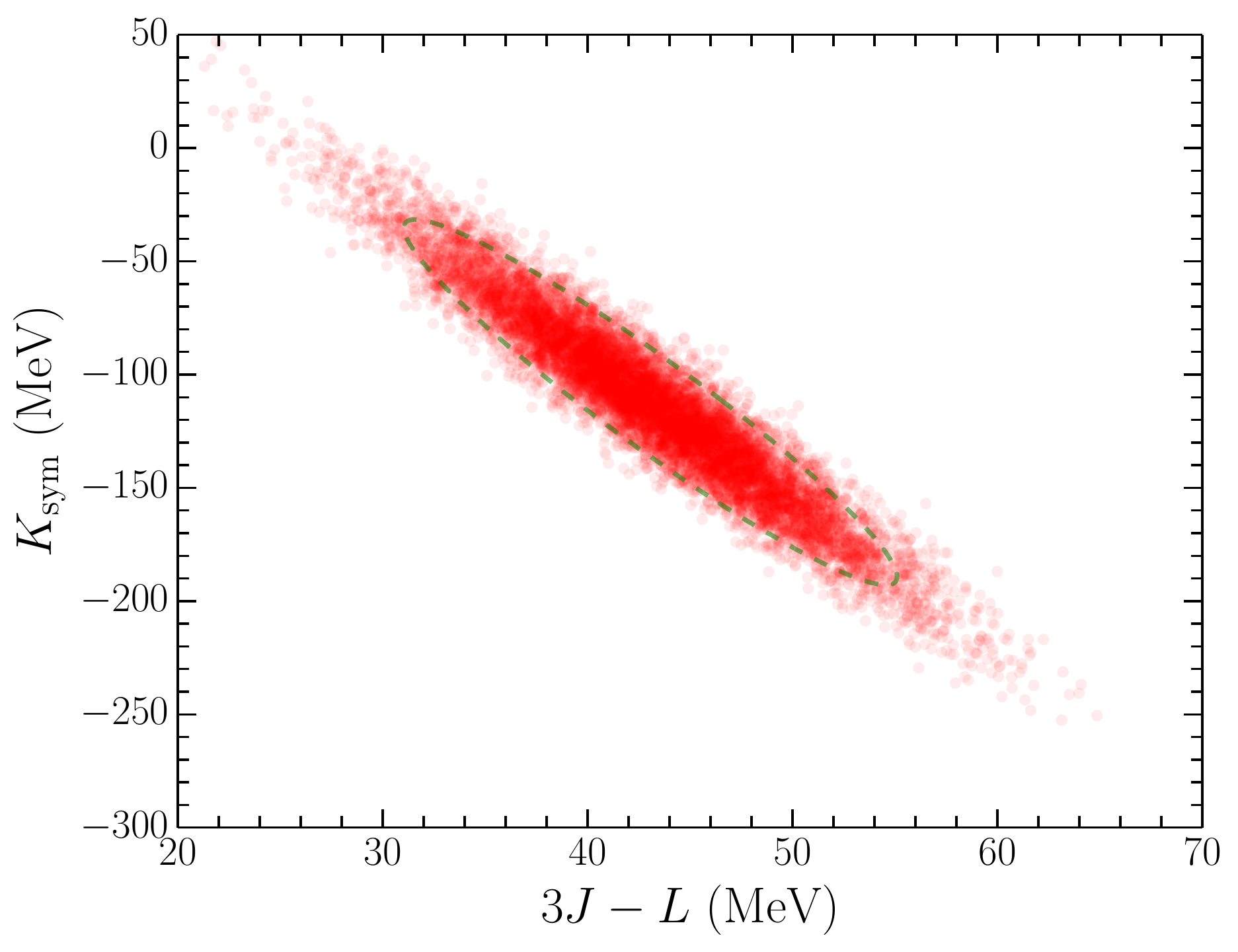}
\caption{Distribution for $K_{\rm sym}$ and $3J-L$ of the energy density functionals generated from the posterior probability distributions in this work. The dashed line denotes the $2\sigma$ correlation ellipse.}
\label{fig:j3ml}
\end{figure}

In Fig.\ \ref{fig:corrsvl} we show the resulting correlation between $J$ and $L$ from the energy density functionals generated by the posterior probability distributions. The closed dashed line denotes the $2\sigma$ correlation ellipse. The angle $\alpha$ between the aphelion-axis and $L$-axis is given as
\begin{equation}
\tan(2\alpha) = \frac{2R_{xy} \sigma_x \sigma_y}{\sigma_x^2 - \sigma_y^2}\,,
\quad
\alpha = - \ang{7.178}.
\end{equation}
Note that we can also see a similar correlation between $J$ and $L$ from the liquid drop model and from the Hartree-Fock approach for nuclear masses \cite{kortelainen10,lattimer13}. As seen in Fig.\,\ref{fig:corrsvl}, our findings for $a_i$ and $b_i$ naturally imply the correlation among nuclear matter properties.
In Fig.\ \ref{fig:j3ml} we plot the posterior symmetry incompressibility $K_{\rm sym}$ against the combination $3J-L$, which were found in \cite{margueron18eob} to be highly correlated. We find that the correlation between the two parameters is $R_{xy}=-0.955$ with $\langle 3J - L \rangle = 43.1\,\mathrm{MeV}$, $\sigma_{3J-L}=6.02$\,MeV, $\langle K_{\mathrm{sym}} \rangle = -112\,\mathrm{MeV}$,  $\sigma_{K_{\mathrm{sym}}}=40.3$\,MeV, and $\alpha = \ang{8.12} $.

\subsection{Neutron star crust}
The inhomogeneous nuclear matter in the crust of a neutron star represents a phase co-existence problem between dense and dilute matter\,\cite{lim17}. The density of the heavy nucleus corresponds to the dense phase while the unbound neutrons correspond to the dilute phase. In the present work we compute the equation of state in the crust of neutron stars using the liquid drop model technique.  The total energy has contributions from the heavy nucleus, unbound neutrons, and electrons:
\begin{equation}
\label{eq:eos}
	\begin{aligned}
		\varepsilon =  & u n_i f_i + \frac{\sigma(x_i)u d}{r_N}
		+ 2\pi (n_i x_i e r_N)^2  u f_d(u) \\
		& + (1-u)n_{no}f_{no} + \varepsilon_e \,,
	\end{aligned}
\end{equation}
where $f_i$ and $f_{no}$ are the nucleonic contributions to the total energy from the heavy nucleus and neutron gas outside, respectively, $n_i$ is the number density of heavy nuclei, $n_{no}$ is the density of the unbound neutron gas, $x_i$ is the proton fraction, $r_N$ is the heavy nucleus radius, and $u$ is the filling factor (the fraction of space taken up by a heavy nucleus
in the Wigner-Seitz cell). The second term $\sigma(x_i)$ in the above equation stands for the surface tension as a function of proton fraction. Finally, $f_d$ is a geometric function describing the Coulomb interaction~\cite{Ravenhall:1983uh} for different dimensions $d$.

The surface tension $\sigma(x_i)$ is computed from the semi-infinite nuclear matter density profile where the dense phase has the proton fraction $x_i$. We adopt the fitting function for the numerical calculation of the surface tension, approximated by
\begin{equation}
\sigma(x) = \sigma_0 
\frac{2\cdot 2^{\alpha} + q}{(1-x)^{-\alpha} + q +  x^{-\alpha}}\,.
\end{equation}
The third term in Eq.\ \eqref{eq:eos} represents the Coulomb energy, with contributions from proton-proton, proton-electron, and electron-electron interactions. The shape function $f_d(u)$ takes into account nuclear pasta phases \cite{lseos} and is an analytic function of the dimension $d$ and volume fraction $u$ of the heavy nucleus in the Wigner-Seitz cell. Because of the nuclear virial theorem, we can obtain a simplified equation for the total energy \cite{lseos}:
\begin{equation}
\varepsilon =   u n_i f_i + \beta \mathcal{D} + (1-u)n_{no}f_{no} + \varepsilon_e \,,
\end{equation}
where $\beta = \left(\frac{243\pi}{5}e^2 x_i^2 n_i^2 \sigma^2 \right)^{1/3}$ and $\mathcal{D}=\mathcal{D}(u)$ is a continuous dimension function. The energy density of electrons is denoted by $\varepsilon_e$. At $T=0$\,MeV, $\mu_e = \frac{\pt \varepsilon_e}{\pt n_e} = \sqrt{m_e^2 + p_{f_e}^2}$\,.

From the constraints on the total baryon number density $n$ and proton fraction $Y_e$ in the cell
\begin{eqnarray}
 n     &=& u n_i+ (1-u)n_{no}\,, \\
 nY_e  &=& u n_i x_i \,,
\end{eqnarray}
we finally have five equations to solve with five unknowns ($u$, $n_i$, $x_i$, $n_{no}$,$Y_e$):
\begin{subequations}
\begin{align}
& p_i + \frac{2\beta \mathcal{D}}{3u} - \beta \mathcal{D}^\prime - p_o = 0 \,, \\
& \mu_{ni} - \frac{2\beta x_i \mathcal{D \sigma^\prime} }{3u n_i x_i \sigma} - \mu_{no}=0 \,,\\ 
& \mu_{ni} - \left(\frac{2\beta \mathcal{D} }{3u n_i x_i} + \frac{2\beta \mathcal{D}\sigma^\prime}{3un_i\sigma} \right)
-\mu_{pi} - \mu_e = 0\,, \\
& n     - u n_i - (1-u)n_{no}=0 \,, \\
& nY_e  - u n_i x_i = 0 \,,
\end{align}
\end{subequations}
where $\sigma^\prime = \frac{\pt \sigma}{\pt x_i}$ and $\mathcal{D}^\prime = \frac{\pt \mathcal{D}}{\pt u}$.

\section{Tidal deformability}
\label{sec:tidal}
The macroscopic structure of the neutron star is computed by solving the Tolman-Oppenheimer-Volkoff (TOV) equations,
\begin{subequations}
	\begin{align}
		\frac{dp}{dr} & = - \frac{(\varepsilon + p)(m +4\pi r^3 p)}{r(r-2m)} \,,\label{eq:dpdr} \\
		\frac{dm}{dr} & = 4\pi r^2 \varepsilon\,,\label{eq:dedr}
	\end{align}
\end{subequations}
where $dp/dr$ describes the pressure change with the distance $r$ from the center. Since the pressure at the center of the neutron star is the highest, the pressure decreases with increasing $r$ according to Eq.\ \eqref{eq:dpdr}. The energy density also decreases as the distance from the center increases. Eq.\ \eqref{eq:dedr} describes how the total gravitational mass of the neutron star increases with the distance. The nuclear equation of state then provides the input for $p$ and $\varepsilon$.

\begin{figure}[t]
	\centering
	\includegraphics[scale=0.54]{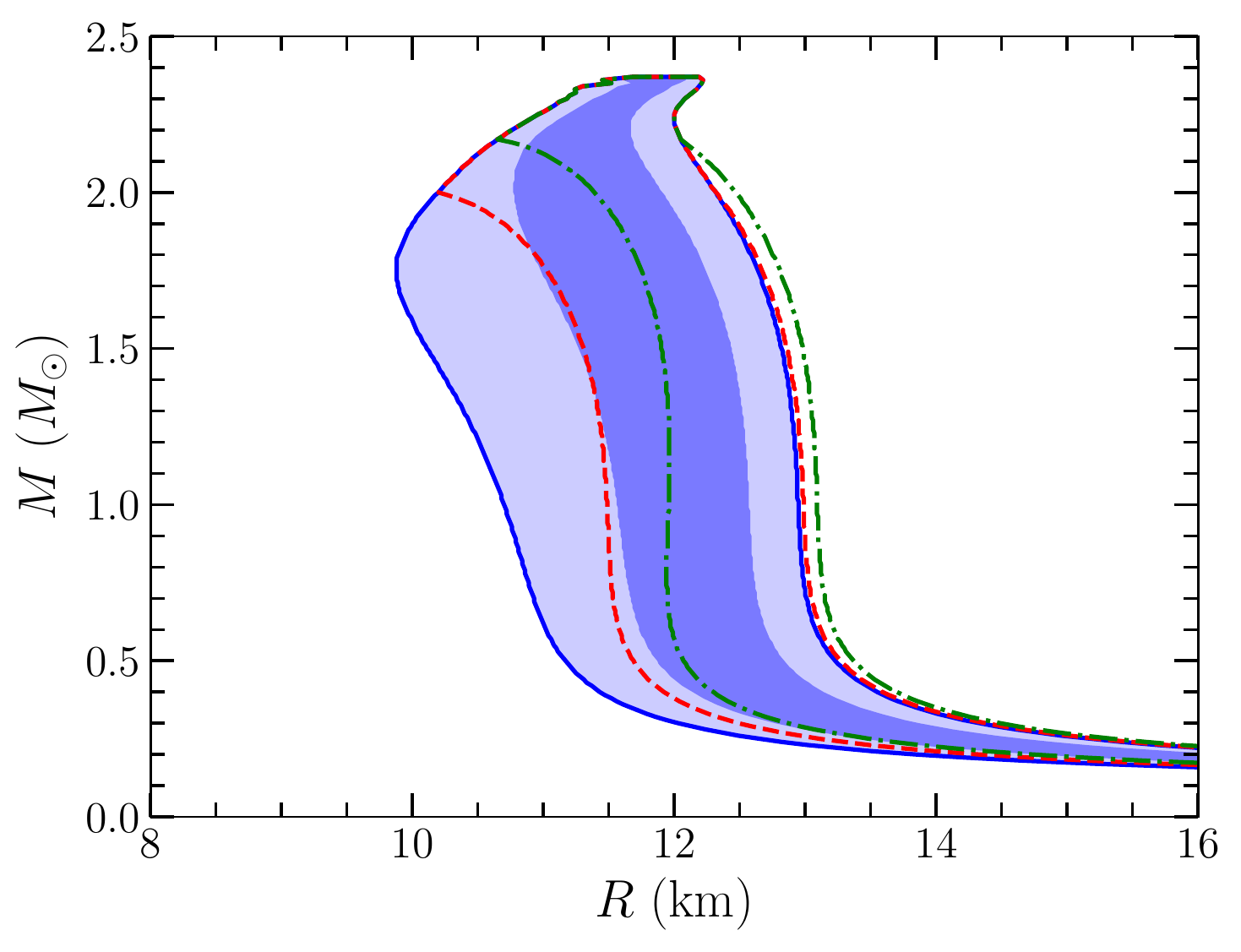}
	\caption{The blue band shows the 95\% (68\%) credibility range for the mass-radius relationship of neutron stars obtained within the present Bayesian modeling of the nuclear equation of state. The red dashed curve area is obtained when we include only those equations of state that produce
		mass of neutron stars greater 2.0\,$\msun$ neutron stars. The green dot-dashed curve indicated 95\% credibility when we include the EOS which 
		can make $2.17\,\msun$ neutron stars.}
	\label{fig:mrrange}
\end{figure}

\label{sec:tidal}
The gravitational wave signal from the late inspiral phase of binary neutron star coalescence is 
connected \cite{hinderer08,read09} to the neutron star equation of state through the dimensionless
tidal deformability
$\Lambda$, which can be determined from the Love number $k_2 = \frac{3}{2}\Lambda \beta^5$
defined through
\begin{eqnarray}
k_2(\beta,y_R) &&= \frac{8}{5}\beta^5 (1-2\beta)^2
\Bigl\{ 2-y_R + 2\beta(y_R-1)\Bigr\} \\ \nonumber
&& \hspace{-.4in}\times  \biggl[ 2\beta
\Bigl\{ 6 -3y_R +3\beta(5y_R-8)\Bigr\} \\ \nonumber
&&\hspace{-.4in} +4\beta^3 \Bigl\{ 13 -11y_R +\beta(3y_R -2)
+ 2\beta^2(1+y_R) \Bigr\} \\ \nonumber
&& \hspace{-.4in}+ 3(1-2\beta)^2 \Bigl\{ 2 - y_R + 2\beta(y_R-1)\Bigr\} \ln(1-2\beta)
\biggr]^{-1},
\end{eqnarray}
where $\beta = M/R$ is the neutron star compactness and $y_R$ is the solution at the neutron 
star surface to the first order differential equation
\begin{equation}
\begin{aligned}
r y^\prime(r)  + y(r)^2 
& + y(r) e^{\lambda(r)}[1 + 4\pi r^2 \{ p(r) - \varepsilon(r) \}] \\
& + r^2 Q(r) =0\,.
\end{aligned}
\label{grav1}
\end{equation}
Here $\varepsilon(r)$ is the energy density and $p(r)$ is the pressure obtained from the
equation of state. In Eq.\ (\ref{grav1}),
$e^{\lambda(r)}$ is the metric function for a spherical star
\begin{equation}
e^{\lambda(r)} = \biggl[ 1 - \frac{2m(r)}{r}\biggr]^{-1}
\end{equation}
and
\begin{equation}
\begin{aligned}
Q(r) = & 4\pi e^{\lambda(r)}
\biggl[5\varepsilon(r) + 9p(r) + \frac{\vare(r) + p(r)}{c_s^2}\biggr] \\
& - 6 \frac{e^{\lambda(r)}}{r^2} 
- 4 \frac{ e^{2\lambda(r)} }{r^4}\Bigr\{m(r) + 4\pi p(r)r^3 \Bigr \}^2\,.
\end{aligned}
\end{equation}
These equations are solved together with the TOV equations for hydrostatic equilibrium to obtain the neutron star mass vs.\ radius relation and tidal deformabilities. The tidal deformability of neutron stars has been studied \cite{Postnikov2010,Hinderer2010,Read2013,Lackey2015,Hotokezaka2016} using many equations of state, including polytropes, realistic nuclear interaction models, and including the presence of quark matter.

\begin{figure}[t]
	\centering
	\includegraphics[scale=0.45]{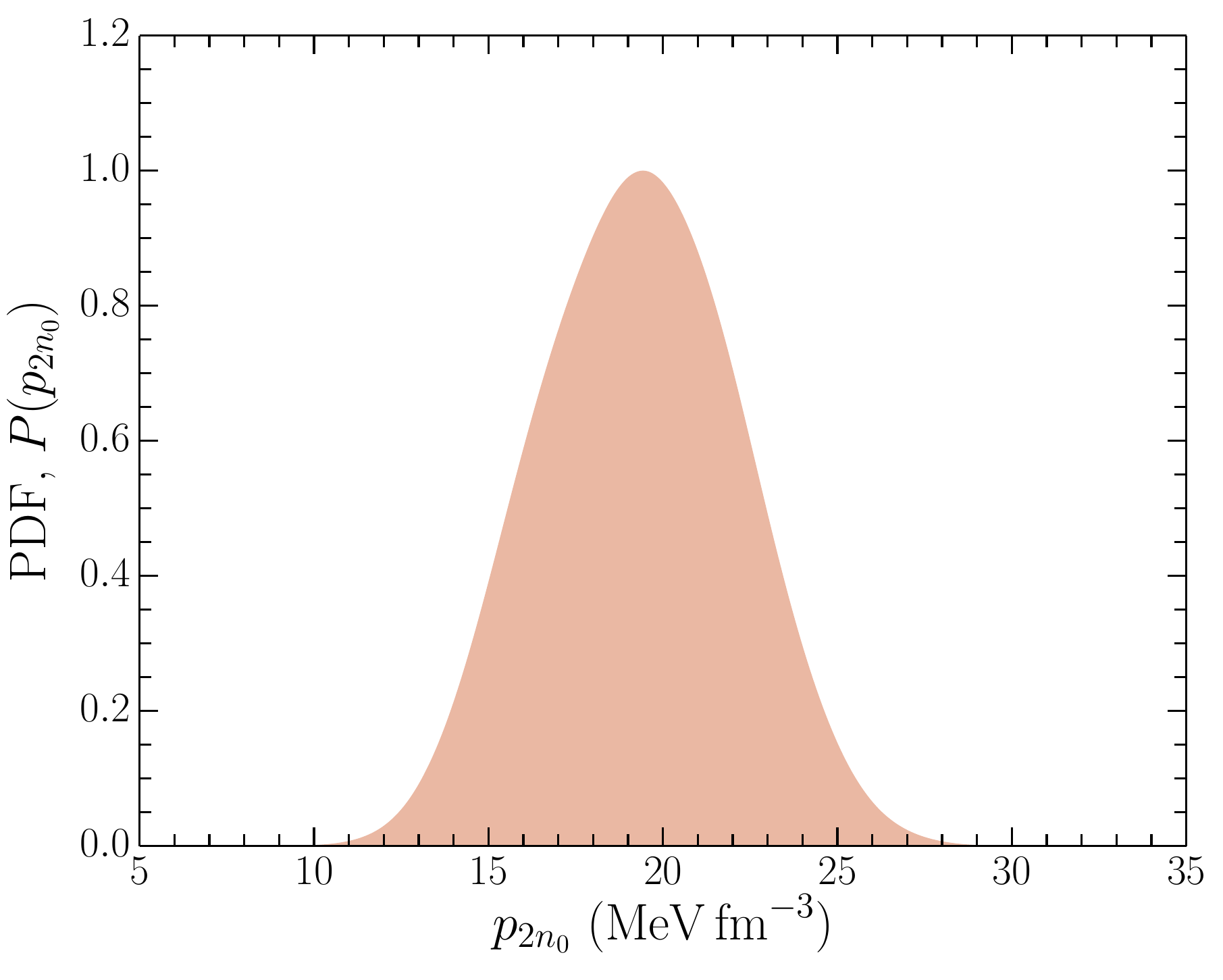}
	\caption{Probability distribution for the pressure of beta equilibrium matter at the density $n=2n_0$ obtained from the present Bayesian modeling of the neutron star equation of state.}
	\label{fig:pdfp2n0}
\end{figure}

\section{Results}
\label{sec:res}

\begin{table}[t]
	\caption{Statistical radius constraints for a given neutron star mass from the 300,000 energy
		density functionals constructed in the present work. The quantity
		$\tilde{R}$ represents the most probable radius for a given mass, while
		$R_{-2\sigma}$ ($R_{+2\sigma}$) and $R_{-\sigma}$ ($R_{+\sigma}$)
		indicates lower (upper) limits of 95\% and 68\% credibility on the radius.}
	\begin{center}
		\begin{tabular}{cccccc}
			\hline
			$M$  &  $R_{-2\sigma}$ & $R_{-\sigma}$  & $\tilde{R}$ 
			& $R_{+\sigma}$  & $R_{+2\sigma}$  \\ 
			\hline
			($M_\odot$) & (km) & (km)  & (km) & (km)  & (km)  \\ 
			\hline
        1.00   &    10.70   &    11.56   &    12.25   &    12.57  &     12.95 \\
        1.10   &    10.61   &    11.53   &    12.25   &    12.56  &     12.94 \\
        1.20   &    10.51   &    11.48   &    12.20   &    12.54  &     12.92 \\
        1.30   &    10.39   &    11.43   &    12.20   &    12.52  &     12.90 \\
        1.40   &    10.26   &    11.36   &    12.15   &    12.48  &     12.87 \\
        1.50   &    10.11   &    11.27   &    12.10   &    12.44  &     12.82 \\
        1.60   &     9.99   &    11.16   &    12.05   &    12.37  &     12.76 \\
        1.70   &     9.89   &    11.04   &    12.00   &    12.29  &     12.68 \\
        1.80   &     9.89   &    10.92   &    11.85   &    12.19  &     12.59 \\
        1.90   &    10.00   &    10.82   &    11.75   &    12.06  &     12.46 \\
        2.00   &    10.20   &    10.77   &    11.55   &    11.91  &     12.32 \\
			\hline
		\end{tabular}
	\end{center}
	\label{mrtab}
\end{table}

\begin{figure}[t]		
	\centering
	\begin{tabular}{c}
		\includegraphics[scale=0.315]{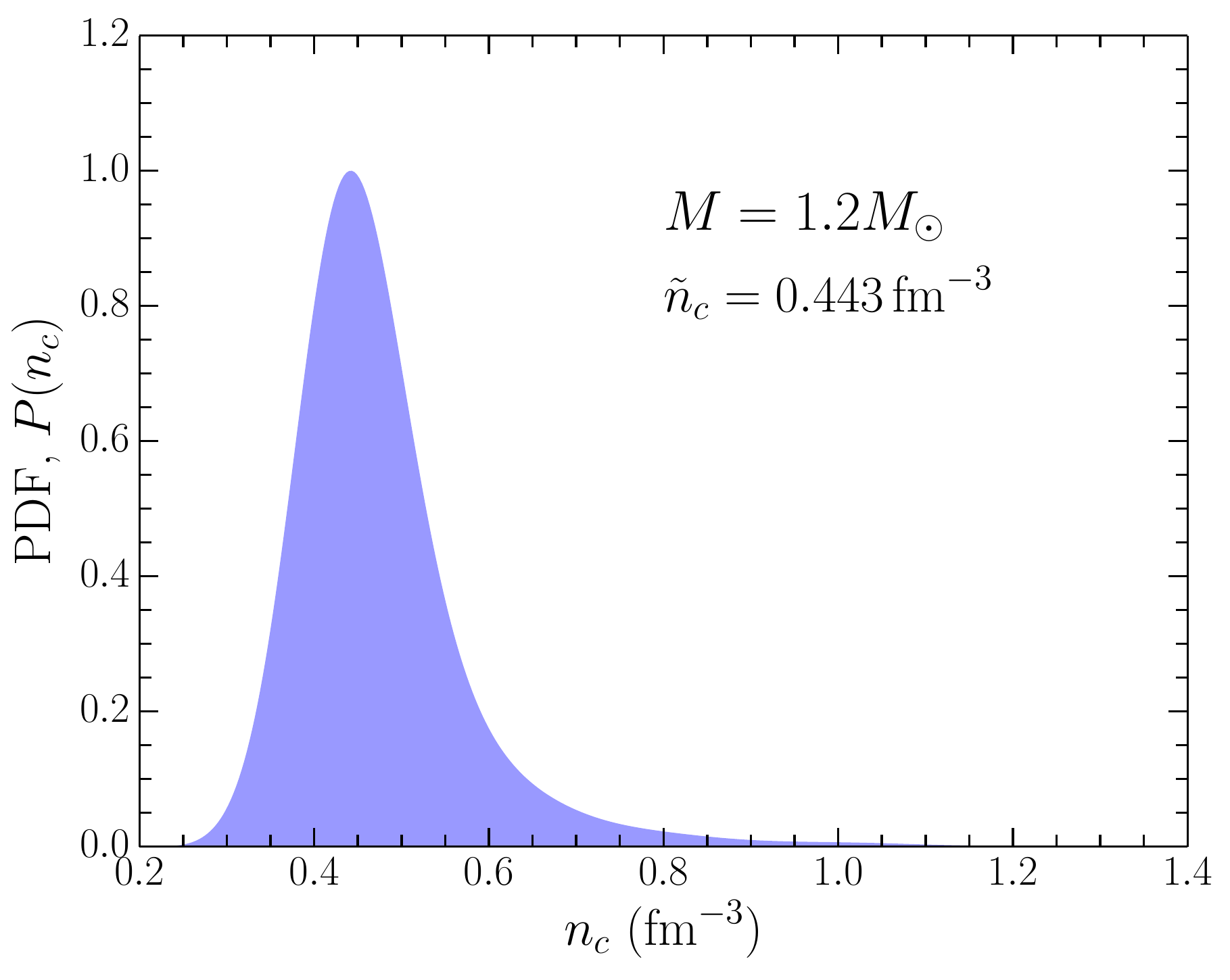} \\
		\includegraphics[scale=0.315]{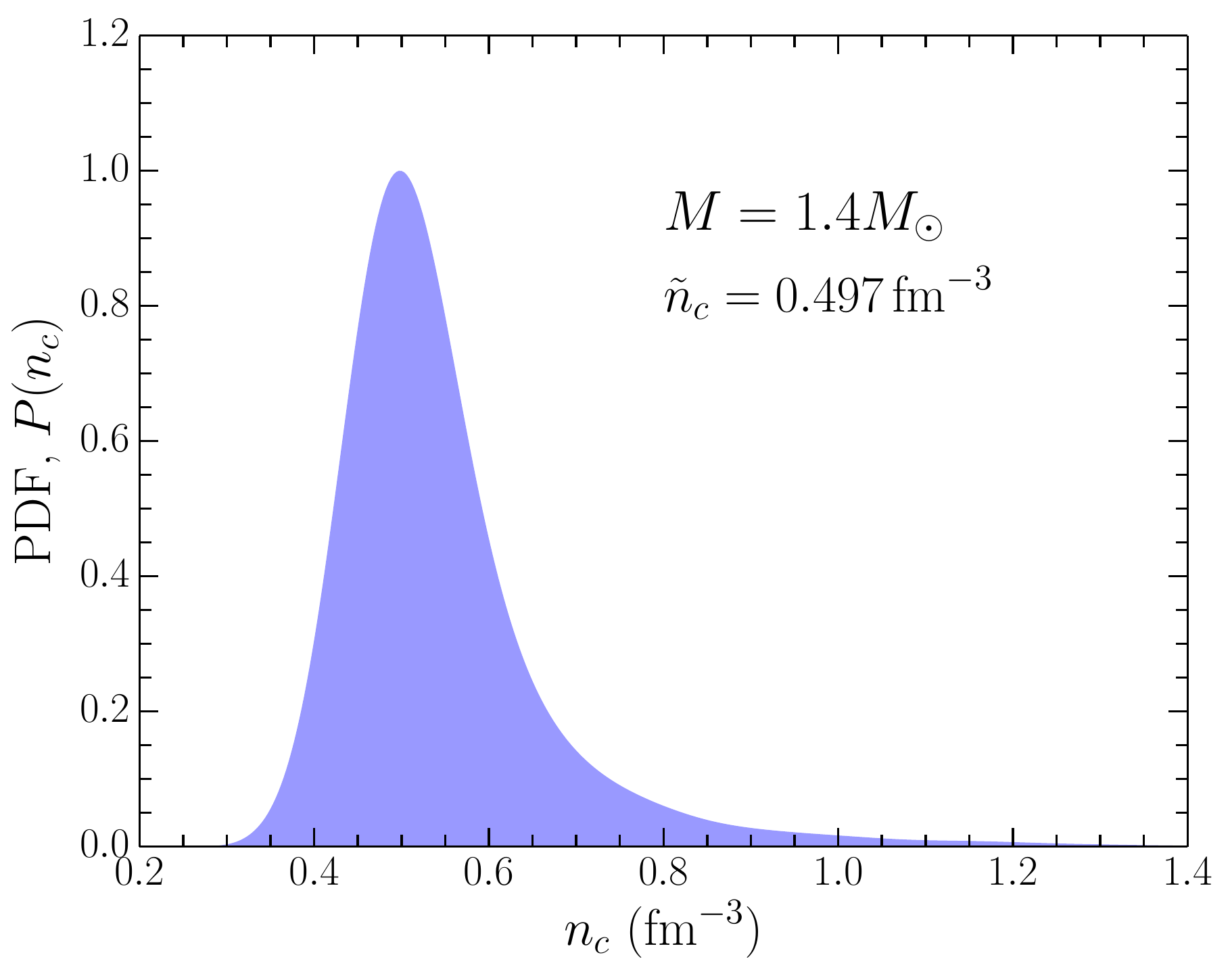} \\
		\includegraphics[scale=0.315]{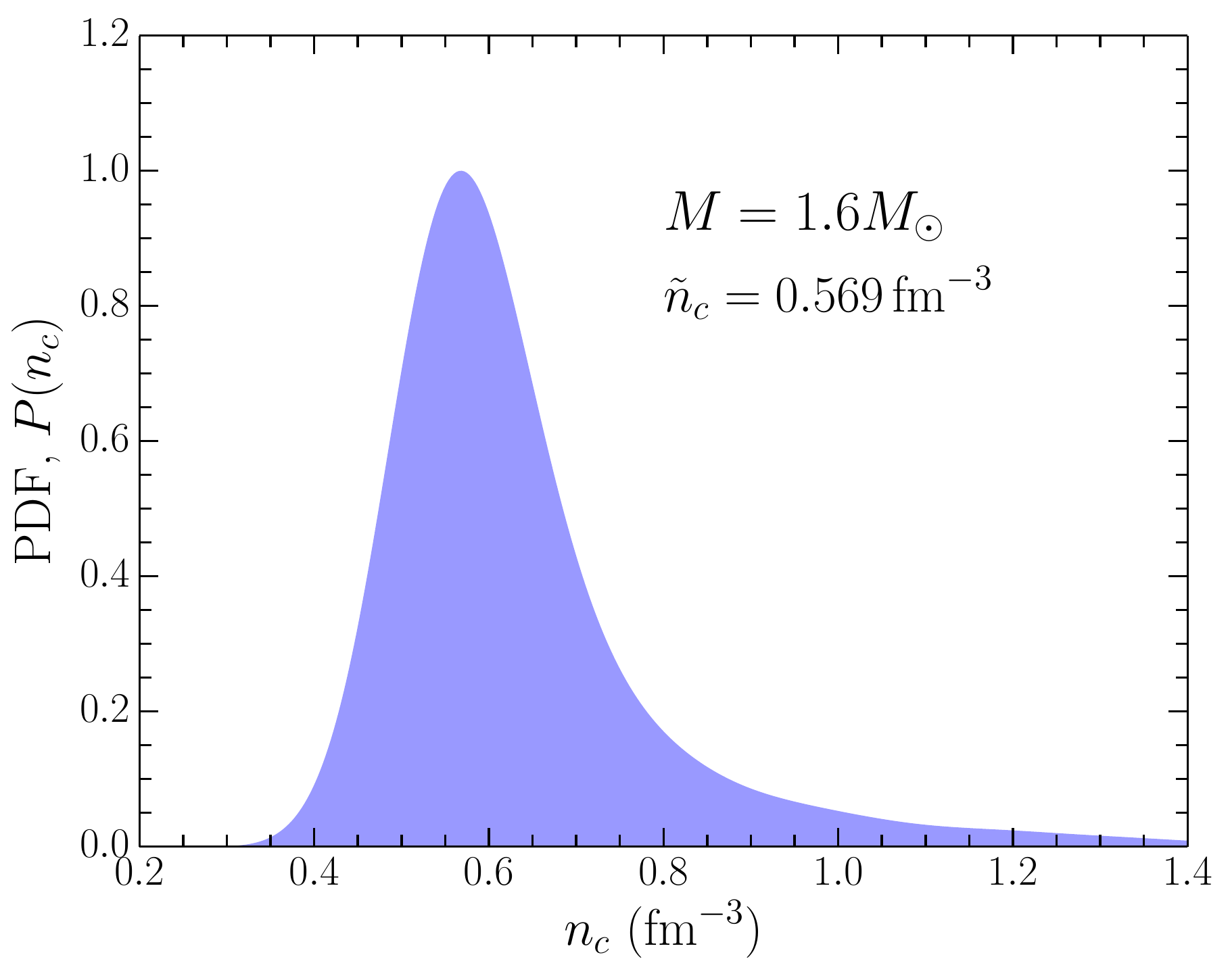} \\
		\includegraphics[scale=0.315]{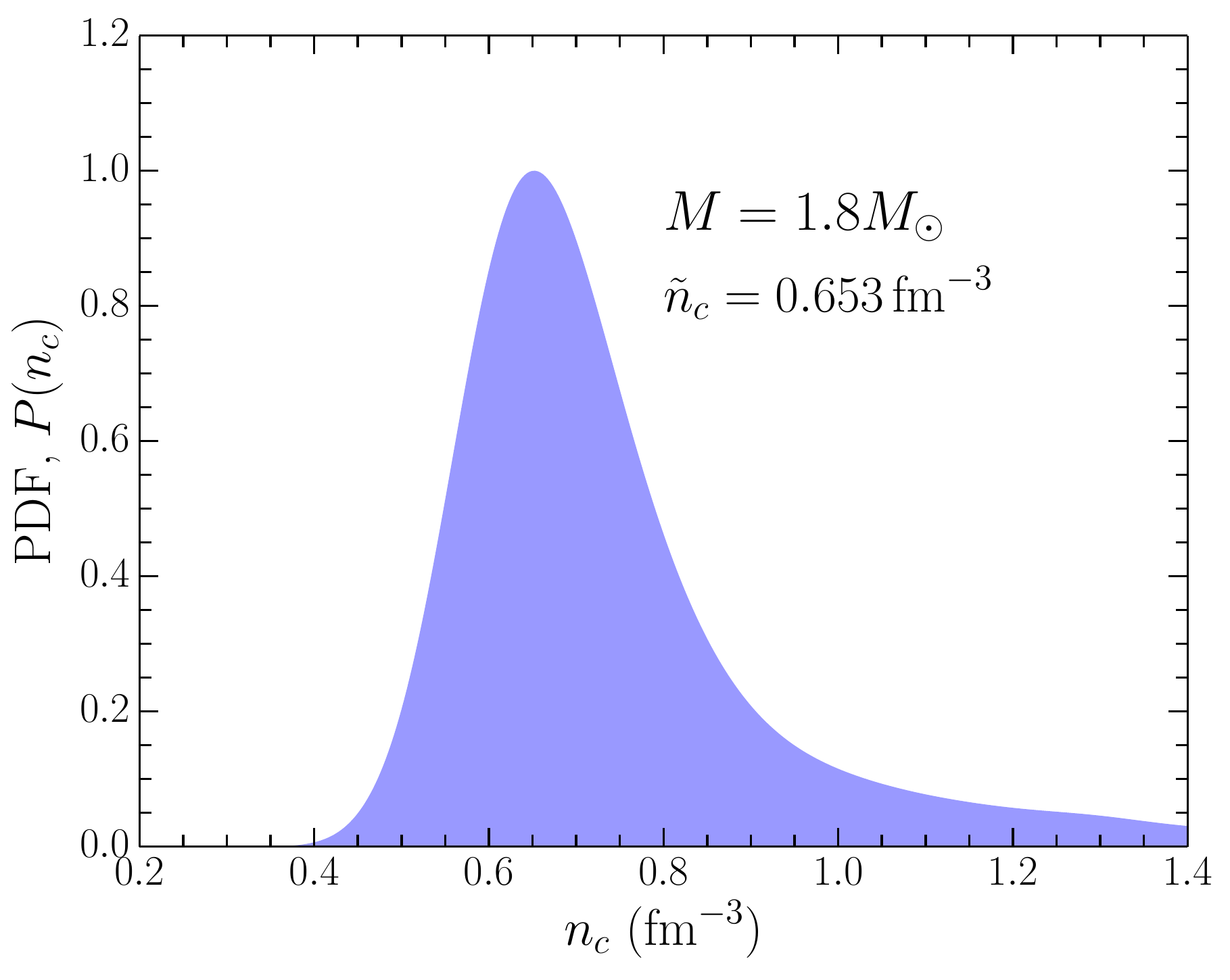}
	\end{tabular}
	\caption{Probability distribution for the neutron star central density $n_c$ obtained in our Bayesian modeling for a series of masses $M = 1.2, 1.4, 1.6, 1.8\,\msun$.}
	\label{fig:massrho}
\end{figure}

In the present work we generate 300,000 neutron star equations of state by sampling from the Bayesian posterior probability distributions for the $a_i$ and $b_i$ parameters. We then compute for each equation of state the mass and radius relation for a cold non-rotating neutron star. As described in the introduction, the particle composition of the neutron star inner core remains highly uncertain and will depend sensitively on the density. For $1.4\,\msun$ neutron stars, the most probable central density from our modeling is around $n_c=0.5\,\mathrm{fm}^{-3}$, which is three times nuclear saturation density. At this high density, nucleon wave functions begin to overlap so that a description in terms of conventional nucleonic degrees of freedom may not be well justified. In addition, the chiral effective field theory expansion is expected to be poorly-behaved. Given the lack of quality experimental constraints on the nuclear equation of state beyond nuclear saturation density, we consider the simplest scenario and naively extrapolate to higher densities using the same functional form as in Eqs. \eqref{eq:edf} and \eqref{eq:fns}. We have confirmed that it is always possible to modify the high-density equation of state ($n>2n_0$) in our generated samples in order to reach a maximum mass of $2.0\,\msun$ while imposing sub-luminal speeds of sound \cite{bedaque15,Constantinou17ztr}. In this work, however, we consider the energy density functional as in Eqs.\ \eqref{eq:edf} and \eqref{eq:fns}, keeping soft equation of states that produce mass-radius relationships with $M_{\rm max} < 2.0\,\msun$ for non-rotating neutron stars.

In Fig.\ \ref{fig:mrrange} we show the 95\% credibility band for the neutron star mass-radius relationship. 
The blue band is the result from all equations of state generated in our Bayesian modeling, the red band shows the results from only those EOSs that can produce $M_{\mathrm{max}} \ge 2.0\,\msun$, and the green band further requires that $M_{\mathrm{max}} \ge 2.17\,\msun$. From the constraint $M_{\mathrm{max}} \ge 2.0\,\msun$, the radius interval for 95\% credibility decreases by around 30\% for typical neutron stars with masses $M \simeq 1.4\,\msun$. For example, the radius credibility interval for a 1.4\,$\msun$ neutron star decreases from $\Delta R_{1.4} = 2.7$\,km to $\Delta R_{1.4} = 1.6$\,km when we impose the additional maximum mass constraint. However, the high-density equation of state remains highly uncertain and a stiffening beyond $n>2n_0$ could repair those models currently rejected from the red band without significantly modifying $R_{1.4}$. Bulk neutron star properties are in fact strongly correlated with the pressure at $n=2n_0$. In Fig.\ \ref{fig:pdfp2n0} we show the probability distribution for the pressure of beta-equilibrium matter at $n=2n_0$ obtained within our Bayesian modeling of the nuclear equation of state. The pressure includes contributions from both nucleons and leptons. We find that the most probable value of the pressure is $\tilde{p}_{2n_0}=19.5\,\mathrm{MeV\,fm}^{-3}$, while the 95\% credibility range for $p_{2n_0}$ is $13.8 \le p_{2n_0} \le 24.9\,\mathrm{MeV\,fm}^{-3} $. Recent work by Abbott \textit{et al}.\,\cite{abbott2018b} re-analyzed data from GW170817 to obtain for the 90\% confidence interval on the pressure $11.235$ $\le p_{2n_0} \le 38.7$ MeV fm$^{-3}$, with central value of pressure $p_{2n_0}=21.8$ MeV fm$^{-3}$. We note that the combined predictions from nuclear theory and experiment obtained in our work lies completely within the range from gravitational wave analyses.

Table \ref{mrtab} shows the statistical distribution for neutron star radii as a function of mass from the energy density functionals constructed in the present work. At the $95\%$ credibility level, the radius of a 1.4\,$\msun$ neutron star is constrained to within $2.7$\,km, having a most probable value of $R_{1.4}=12.0$\,km. We observe that the most probable radius and the $95\%$ credibility region for the radius do not change rapidly in the mass range between $1.0$ and $1.5\,\msun$. Compared with the previous mass and radius range from X-ray burst data analysis, our results from the energy density functionals based on chiral effective field theory and nuclear experiments give similar results to Steiner \textit{et al.} \cite{SLB2010,lattimer14,SLB2016}. Current analysis of tidal deformability constraints from GW170817 \cite{abbott2018b} give for the radius of $M \sim 1.4\,\msun$ neutron stars the value $R=11.9^{+1.4}_{-1.4}$\,km. This is in close agreement with our credibility interval for the mass and radius from nuclear modeling.

A key quantity associated with the possibility of phase transitions in neutron stars is the central density of the inner core. In Fig.\ \ref{fig:massrho} we show the resulting statistical distribution of central densities for neutron star masses $M = 1.2, 1.4, 1.6, 1.8\,\msun$ from our Bayesian modeling of the nuclear equation of state. Generically, the most probable central density increases with the mass of the neutron star. The $68\% (\pm \sigma)$ and $95\% (\pm 2\sigma)$ credibility intervals for the central density widen as the mass of the neutron star increases. This implies that the uncertainty increases as the baryon number density increases.  We see that the lightest neutron stars (with $M \simeq 1.2 \msun$) are predicted on average to have central densities less than about three times normal nuclear matter density $n_0$. However, for heavier neutron stars with mass $M \simeq 1.8 \msun$, the most probable central density is greater than four times nuclear saturation density and the distribution extends significantly higher to $6-7\,n_0$, where a description in terms of well defined nucleonic degrees of freedom would be questionable due to the fact that the nucleons are strongly overlapping. Nevertheless, as a minimal scenario we presently assume no phase transitions as well as the absence of higher-order powers of the Fermi momentum in the nuclear energy density functional.

\begin{figure}[t]
\centering
\includegraphics[scale=0.56]{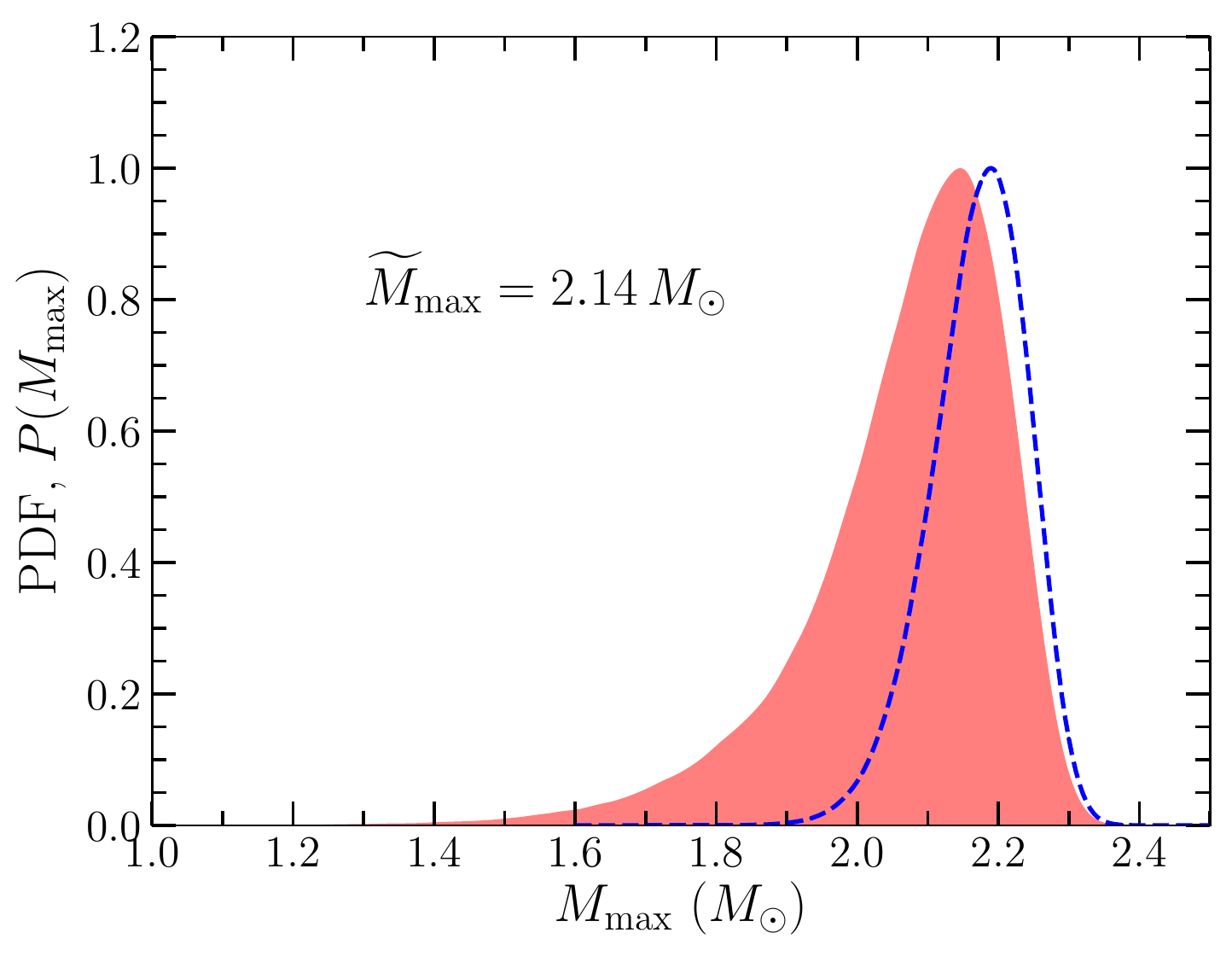}
\caption{Probability distribution (red) for the maximum neutron star mass derived from our Bayesian modeling of the nuclear equation of state including constraints from nuclear theory and experiment. The blue dashed line represents the probability distribution when we include in our posterior the $M=2.17^{+0.11}_{-0.10}\,\msun$ observation \cite{Cromartie19}.}
\label{fig:pdfnsmax}
\end{figure}

Most of the equations of state derived from the Bayesian posterior probability distributions produce a mass-radius relation with a maximum mass $M_{\rm max} > 2\,\msun$. In Fig.\ \ref{fig:pdfnsmax} we show the probability distribution for this maximum mass, which has a most probable value of $2.14\,\msun$ and an average value of $2.04$ $\msun$. From observational data associated with the electromagnetic counterpart to GW170817, numerous authors have argued that the most likely post-merger object was a relatively long-lived hypermassive neutron star. Such a scenario rules out both very soft and very stiff equations of state, the former would have led to a prompt collapse and black hole formation, while the latter would have resulted in a very long-lived supramassive neutron star. These arguments \cite{bauswein13,lawrence15,fryer15} have been used \cite{margalit17,shibata17,rezzolla18,ruiz18} to give the first observational upper bounds on the maximum neutron star mass $M_{\rm max} < 2.2 - 2.3\,\msun$. In our modeling approximately 75\% of the equations of state give a maximum neutron star mass greater than $2.0\,\msun$, and very few produce maximum masses greater than $M_{\rm max} = 2.3\,\msun$ as seen in Fig.\ \ref{fig:pdfnsmax}.
We have also included constraints from the observed $M=2.17^{+0.11}_{-0.10}\,\msun$ neutron star from Ref.\ \cite{Cromartie19}. Since the mass distribution has a large variance, we impose the full posterior in our Bayesian analysis (see e.g., Refs.\ \cite{Alvarez16,Miller19a}). The modified maximum mass distribution is shown as the blue band in Fig.\ \ref{fig:pdfnsmax}.

\begin{figure}
\centering
\includegraphics[scale=0.60]{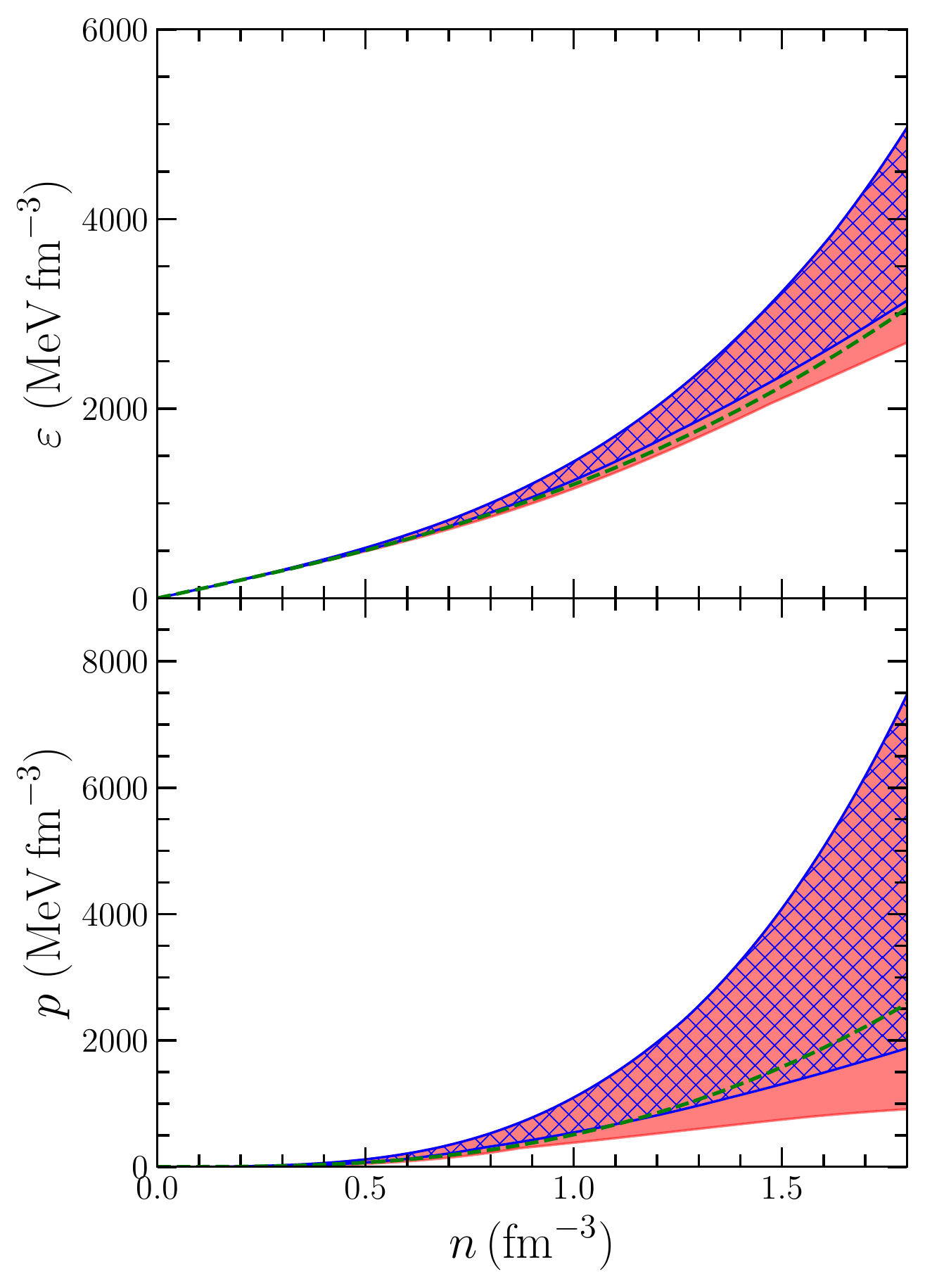}
\caption{Neutron star energy density and pressure as a function of baryon number density. The red bands show the
	energy density and pressure which produce maximum neutron star masses greater than $2.0\,\msun$, while
	the blue hatched regions produce maximum neutron star masses greater than $2.17\,\msun$. The dashed green 
	curves denote results from the SLy4 Skyrme effective interaction.}
\label{fig:epminmax}
\end{figure}

In Fig.\ \ref{fig:epminmax} we show the neutron star energy density and pressure as a function of the total baryon number density.
The red (blue) band shows the range of equations of state when we impose that the maximum mass be greater than $2.0\,\msun$ ($2.17\,\msun$). The EOS based on the SLy4 Skyrme effective interaction is added for comparison. Note that SLy4 gives a maximum neutron star mass of $2.06\,\msun$. Since we sample 300,000 equations of state, these bands may be useful to test whether other models can reach current maximum neutron star mass constraints.

\begin{figure}[t]
	\centering
	\includegraphics[scale=0.56]{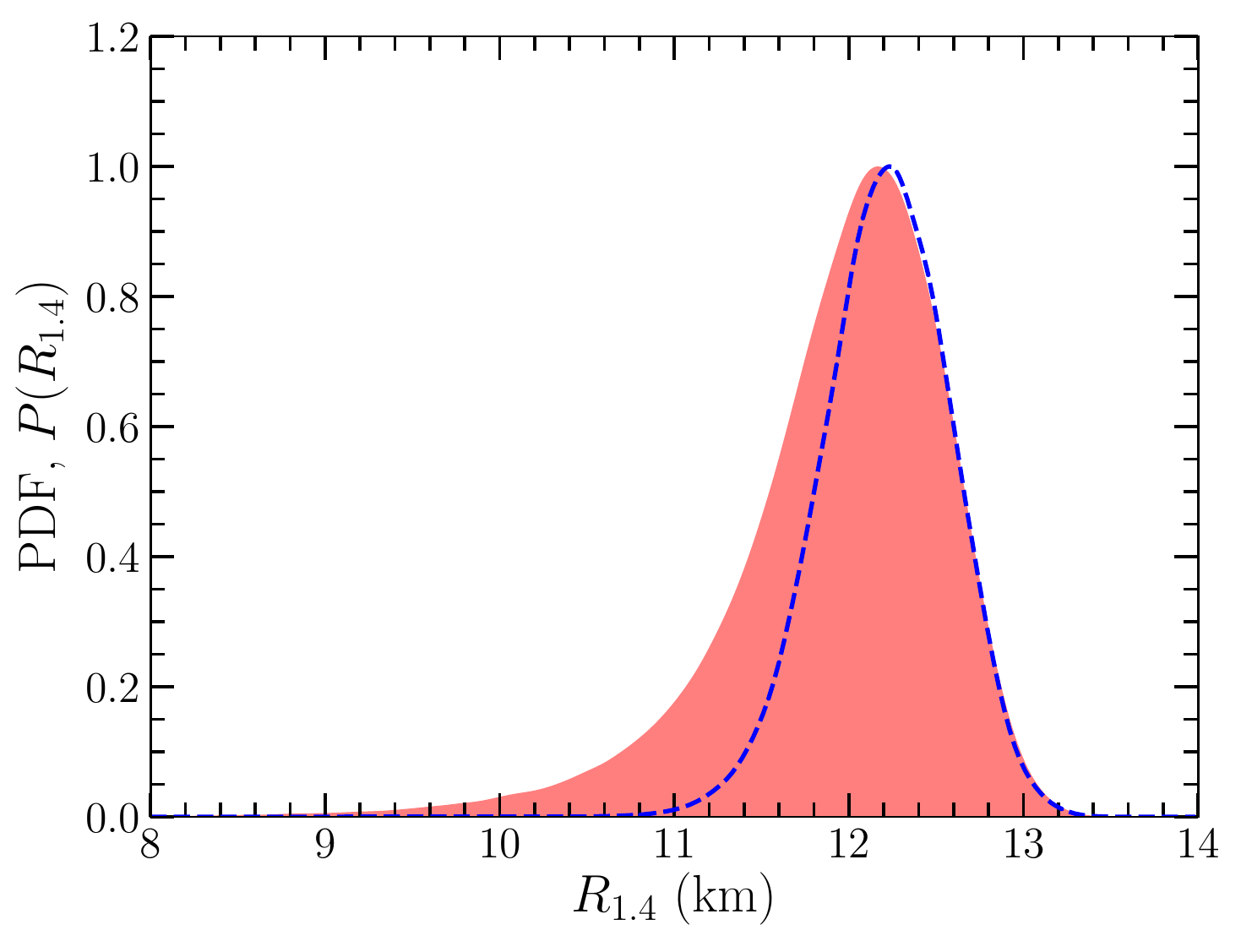}
	\caption{Probability distribution (red) for the radius of a $1.4\,\msun$ neutron star from the Bayesian analysis in the present work.
		The blue dashed line represents the probability distribution when we include in our posterior for $R_{1.4}$ 
		the $M=2.17^{+0.11}_{-0.10}\,\msun$ observation \cite{Cromartie19}.}
	\label{fig:pdfmr14}
\end{figure}

\begin{figure}
\centering
\includegraphics[scale=0.56]{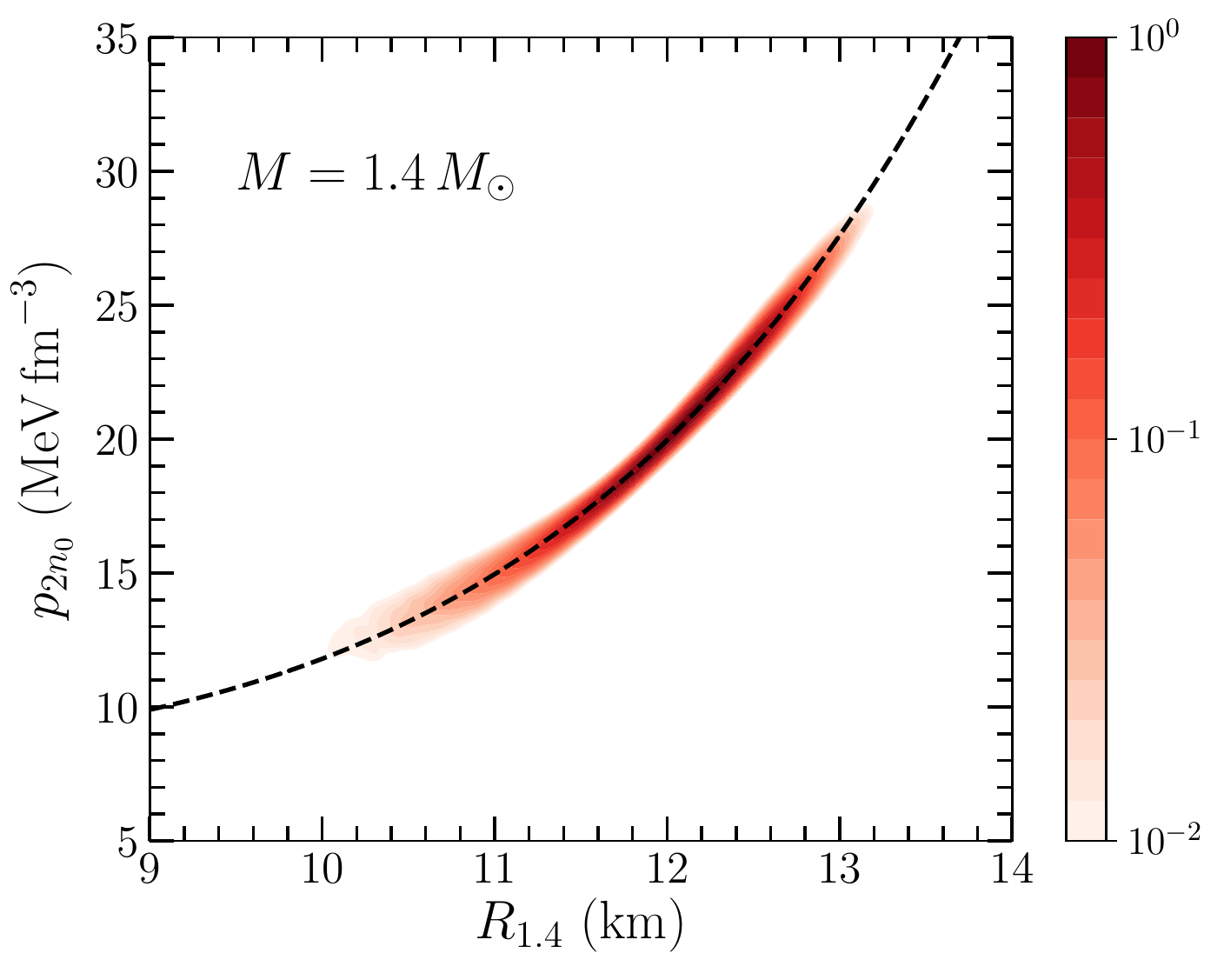}
\caption{Contour plot of the pressure of beta-equilibrium matter at the density $n=2n_0$ and radius of a 1.4\,$\msun$ neutron star. 
	The dashed curve represents the correlation in Eq.\,\eqref{eq:p2n0}.}
\label{fig:rp2n0dist}
\end{figure}

In Fig.\ \ref{fig:pdfmr14} we show the probability distribution for the radius of a $1.4\,\msun$ neutron star obtained within the current modeling of the equation of state. The most probable value of the radius is $R_{1.4}=12\,\mathrm{km}$, but there is a large asymmetry about this central value. In particular, the distribution extends to small radii close to $R_{1.4} \simeq 10$\,km for the softest equations of state generated within our Bayesian analysis. 
Although such models typically produce maximum neutron star masses $M_{\rm max} < 2\,\msun$, as mentioned earlier it is possible to modify the high-density equation of state ($n>2n_0$) to meet this astrophysical constraint while not changing significantly the radius of a $1.4\,\msun$ neutron star. 
Only when we consider EOSs consistent with the $2.17^{+0.11}_{-0.10}\,\msun$ constraint do the radii significantly shift to higher values. In this case $\tilde R = 12.21$\,km, $R_{-2\sigma} = 11.43$\,km, $R_{-\sigma} = 11.84$\,km, $R_{+\sigma} = 12.56$\,km, and $R_{+2\sigma} = 12.88$\,km.
As further evidence that bulk neutron star properties such as the radius and tidal deformability are strongly correlated \cite{lattimer01,lim18a,tsang19} with the pressure of beta-equilibrium matter at the density $n=2n_0$, we show in Fig.\ \ref{fig:rp2n0dist} the probability distribution for the pressure and radius. The dashed curve represents the phenomenological relationship
\begin{equation}
\label{eq:p2n0}
p_{2n_0}= p_i + p_o\left(\frac{R_{1.4}}{12\,\mathrm{km}} \right)^{\alpha}\,,
\end{equation}
where $p_i = 7.63\,\mathrm{MeV\,fm}^{-3}$, $p_o=12.34\,\mathrm{MeV\,fm}^{-3}$, $\alpha=6$,
and the correlation coefficient is $r_{xy}=0.995$.

From the inferred tidal deformability bounds for a $1.4\,\msun$ neutron star from GW170817, together with the most conservative modeling of the equation of state, it has been demonstrated \cite{Fattoyev18,annala18,most18,tews2018gw} that the maximum radius for a $1.4\,\msun$ neutron star is given by $R_{1.4} \lesssim 13.6$\,km. In Fig.\ \ref{fig:pdfmr14} we find that the equations of state employed in the present work are apparently more constrained and typically generate radii $R_{1.4} < 13.0$\,km. This is partly a result of our limited treatment of the high-density equation of state. With additional astrophysical constraints, such as simultaneous mass and radius measurements from the NICER mission, additional tidal deformability bounds from gravitational wave observations, and possibly a measurement of the moment of inertia of pulsar J0737-3039A, the probability distribution for the neutron star mass-radius relation can be narrowed by extending the present Bayesian analysis \cite{raithel17}.

\begin{table}[b]
	\begin{center}
	\caption{Statistical constraints on the neutron star tidal deformability for a given mass from
		the 300,000 energy density functionals constructed in the present work. The quantity
		$\Lambda_c$ represents the most probable value of $\Lambda$ for a given neutron star mass,
		while $\Lambda_{-2\sigma}$ ($\Lambda_{+2\sigma}$) and 
		$\Lambda_{-\sigma}$ ($\Lambda_{+\sigma}$) indicate the lower (upper) limits of 95\% and 68\% 
		credibility respectively.}
	\begin{tabular}{cccccc}
		\hline
		$M$ ($M_\odot$) &  $\Lambda_{-2\sigma}$ & $\Lambda_{-\sigma}$  & $\Lambda_c$     
		& $\Lambda_{+\sigma}$  & $\Lambda_{+2\sigma}$  \\ 
		\hline
        1.0 &    1240   &  2030     &  2800   &    3310   &   3880 \\
        1.1 &     673   &  1170     &  1590   &   1960    &   2300  \\
        1.2 &     372    &  688      &  976    &  1190     &   1410 \\
        1.3 &     208   &   412     &  604    &   738     &  879 \\
        1.4 &     116   &   249     &  379    &   465     &  557 \\
        1.5 &      65 &     150 &    240  &     296 &     357  \\
        1.6 &      37 &      91 &  150  &     189 &    230  \\
        1.7 &      21 &      54 &   93  &     121 &     148  \\
        1.8 &      13 &      32 &   58  &      76 &      95  \\
        1.9 &       9 &      20 &   34  &      48 &      61 \\
        2.0 &       7 &      12 &    19 &      29 &      38 \\
		\hline
	\end{tabular}
	\end{center}
	\label{tidtab}
\end{table}

\begin{figure}[t]
\centering
\includegraphics[scale=0.56]{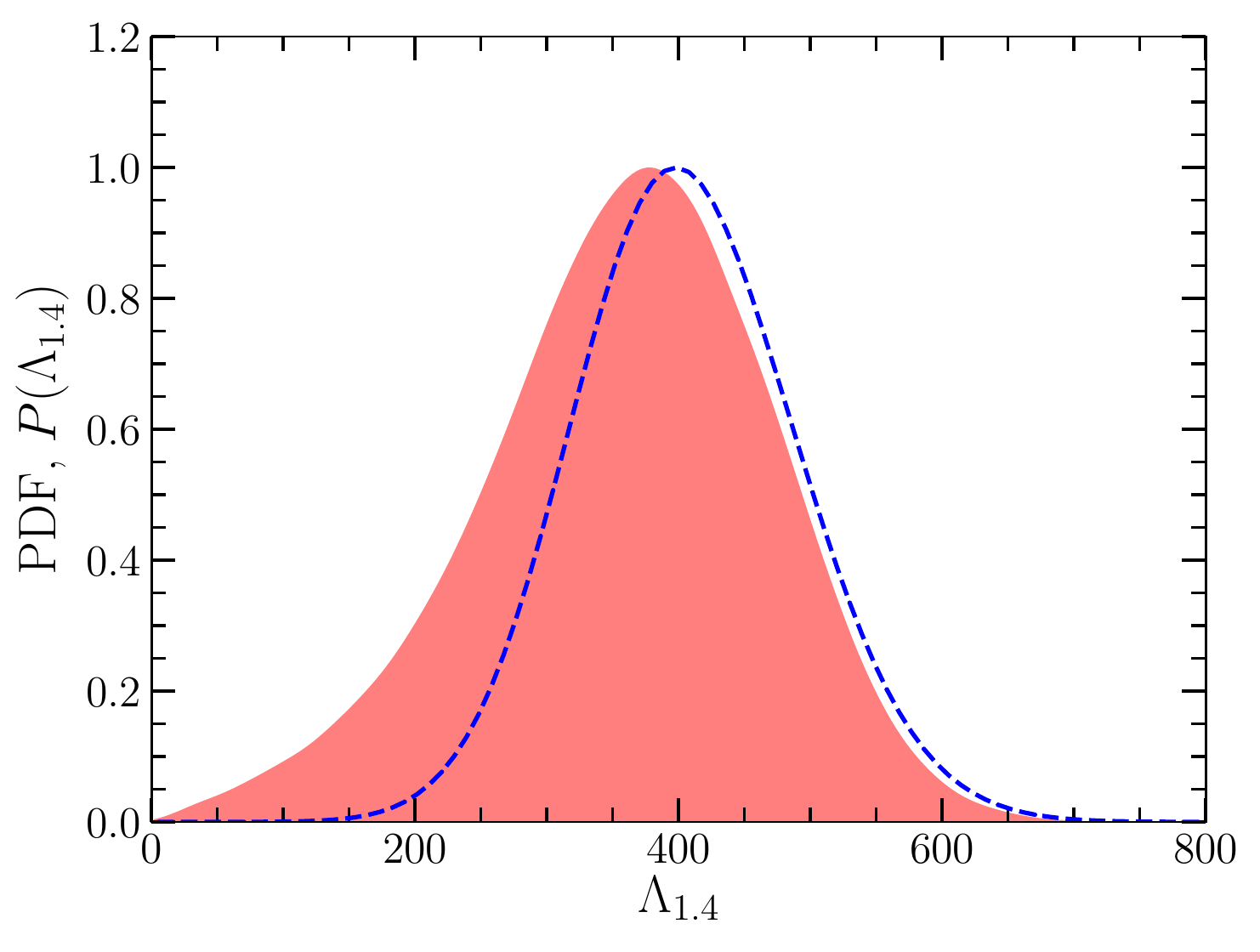}
\caption{Probability distribution (red) for the tidal deformability of a $1.4\,\msun$ neutron star obtained from the 300,000 equations of state generated from the posterior distribution in our Bayesian modeling. The blue dashed line represents the probability distribution when we include in our posterior for $\Lambda_{1.4}$ the $M=2.17^{+0.11}_{-0.10}\,\msun$ observation \cite{Cromartie19}.}
\label{fig:pdfmlb14}
\end{figure}

In Table \ref{tidtab} we show the statistical distribution of tidal deformabilities for a given neutron star mass, ranging from $M=1.0 - 2.0\,\msun$. The tidal deformability rapidly decreases as the mass of the neutron star increases. It was shown \cite{annala18} that the tidal deformability is proportional to $R^{7.5}$, and we confirm that our energy density functional formalism constrained by nuclear theory and experiment also follows such a relation. This suggests that the tidal deformability would have an algebraic relation with the mass of neutron stars. We find
\begin{equation}
\Lambda(M) = \Lambda_{1.4} \left( \frac{1.4\msun}{M}\right)^{a + b M/M_\odot}\,,
\end{equation}
where for our EDFs, ($a=3.76$, $b=2.10$) for the central value of $\Lambda_{50\%}$, 
($a=3.97$, $b=1.73$) for $\Lambda_{+2\sigma}$,
and ($a=6.12$, $b=1.17$) for $\Lambda_{-2\sigma}$. 
Compared with numerical calculations, this fitting function gives a relative error less than $5\%$ when $\Lambda \ge 100$. 

Assuming a common equation of state for the two merging neutron stars, the LIGO collaboration re-analyzed GW170817 and found a tighter bound of $\Lambda_{1.4} = 190^{+390}_{-120}$ at the 90\% confidence level \cite{abbott2018b}. From Table \ref{tidtab} we see that this value is consistent with the present modeling and does not impose strong constraints on our equations of state. 
In Fig.\ \ref{fig:pdfmlb14} we show the probability distribution for the tidal deformability of a $1.4\,\msun$ neutron star based on our Bayesian modeling of the nuclear equation of state. The most probable value of the distribution occurs at $\Lambda_{1.4}=379$, and there is very little probability for tidal deformabilities extending beyond $\Lambda_{1.4}=600$. However, the distribution extends to quite low values of $\Lambda_{1.4}$ for the softest equations of state. Including also the recent observation of a $M=2.17^{+0.11}_{-0.10}\,\msun$ neutron star into the final posterior probability distribution, we show as the blue curve in Fig.\ \ref{fig:pdfmlb14} the
resulting distribution for the tidal deformability of a $1.4\,\msun$ neutron star. We see that the probability to obtain tidal deformabilities with $\Lambda_{1.4} < 200$ become highly unlikely.

\begin{figure}[t]
\centering
\includegraphics[scale=0.45]{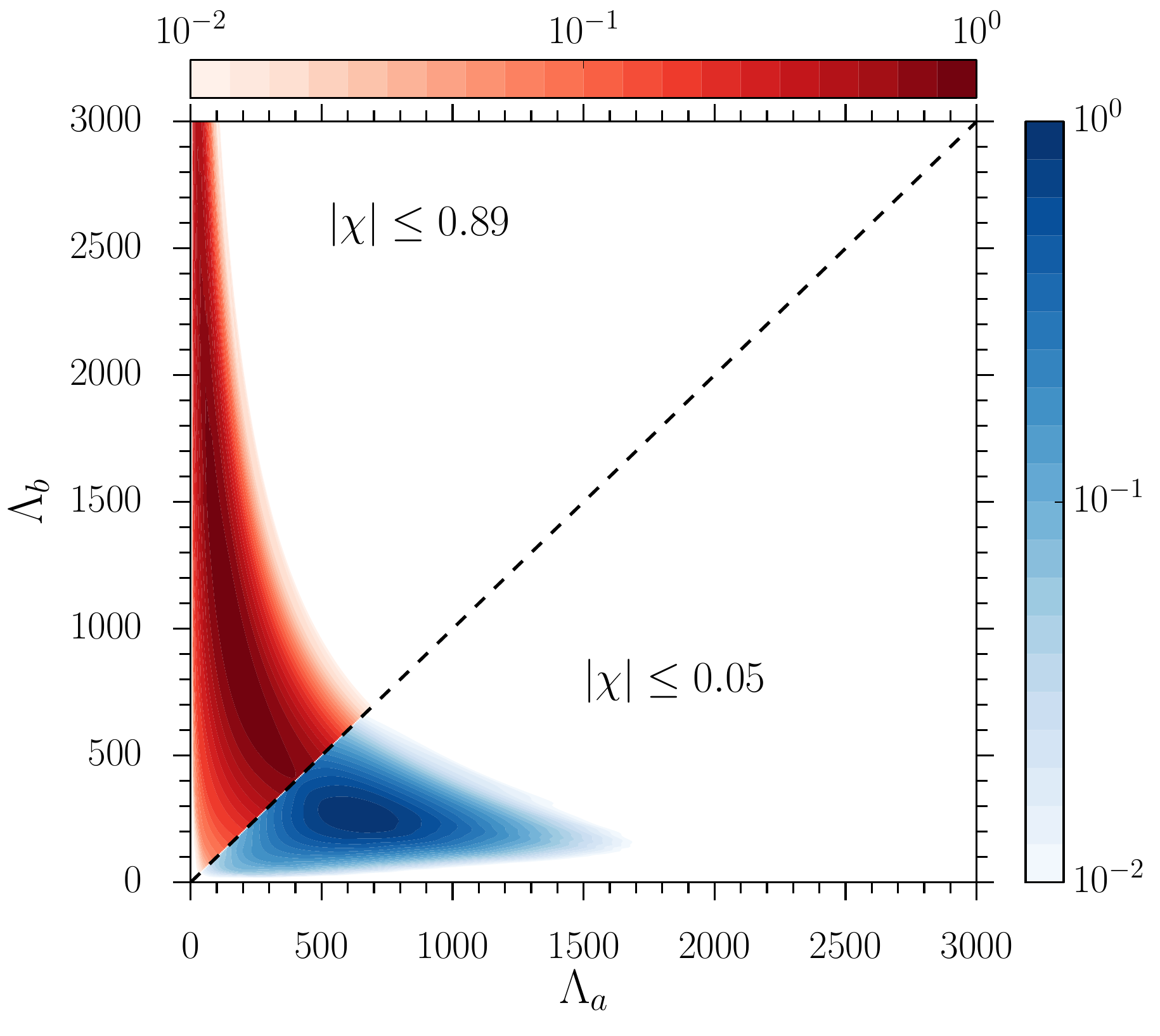}
\caption{Joint probability distribution for $\Lambda_1$ (more massive companion) and $\Lambda_2$ (less massive companion) associated with the two neutron stars in the compact binary of GW170817. The upper left half of the figure corresponds to the high-spins prior mass distributions in Ref.\ \cite{abbott17a}, while the lower right half of the figure corresponds to the low-spins prior mass distributions. In the top-left half, $\Lambda_a = \Lambda_1$ and $\Lambda_b = \Lambda_2$, while in the bottom-right half $\Lambda_a = \Lambda_2$ and $\Lambda_b = \Lambda_1$.}
\label{fig:lamb12}
\end{figure}

For the specific case of GW170817 we show in Fig.\ \ref{fig:lamb12} the joint probability distribution for $\Lambda_1$ and $\Lambda_2$ computed within our Bayesian modeling of the equation of state for a chirp mass
\begin{equation}
{\cal M} = \frac{(m_1 m_2)^{3/5}}{(m_1+m_2)^{1/5}} = 1.188\,\msun.
\end{equation}
The individual probability distributions for $m_1$ (heavier neutron star) and $m_2$ (lighter neutron star) are taken from Ref.\ \cite{abbott17a}, and once the component masses have been sampled, we compute the associated tidal deformabilities from our 300,000 equations of state. In our calculations, both neutron stars are assumed to be governed by the same equation of state. Results for the tidal deformability assuming the low-spin prior ($|\chi| \leq 0.05$) are shown in the bottom-right half of Fig.\ \ref{fig:lamb12} for the case $\Lambda_a = \Lambda_2$ and $\Lambda_b = \Lambda_1$, while results assuming the high-spin prior ($|\chi| \leq 0.89$) are shown in the top-left half of the figure for the case $\Lambda_a = \Lambda_1$ and $\Lambda_b = \Lambda_2$. In the high-spin scenario, the uncertainties in $\Lambda_1$  and $\Lambda_2$ are much larger than that of the low-spin scenario because the mass range of the two neutron stars is much larger in the former. For the low-spins prior, the high-mass neutron star ($m_1,\Lambda_1$) probability distribution peaks at $\Lambda = 258$ while the low-mass neutron star distribution peaks at $\Lambda = 628$. These results for the tidal deformabilities of the neutron stars in GW170817 are consistent with the recent analysis in Ref.\ \cite{de2018a}, though our distributions peak at larger values of $\Lambda$.

Finally, we consider possible correlations between the neutron star tidal deformability $\Lambda$ and the neutron skin thicknesses $\Delta R_{np}$ of neutron-rich nuclei, especially $^{208}$Pb, which has recently been investigated in Ref.\ \cite{Fattoyev18}. It is well known \cite{Roca2011} that the neutron skin thickness is highly correlated with the nuclear symmetry energy slope parameter $L$ in mean field theory calculations based on Skyrme energy density functionals or relativistic mean field (RMF) models. An accurate measurement of the neutron skin thickness may therefore well constrain $L$ and the nuclear equation of state beyond nuclear saturation density. At the same time $L$ is correlated with the neutron star radius and also the tidal deformability. Experimentally, the neutron skin thickness of $^{208}$Pb has been investigated by electric dipole response \cite{Tammi2011}, exotic atoms \cite{FG2007}, hadron scattering \cite{Zenihiro2010}, coherent pion photoproduction \cite{Tarbert2014}, the PREX experiment at JLab \cite{Abrahamyan2012,Horowitz2012}, and the combination of neutron star observations and chiral effective field theory \cite{PhysRevLett.105.161102}.

\begin{figure}[t]
\centering
\includegraphics[scale=0.43]{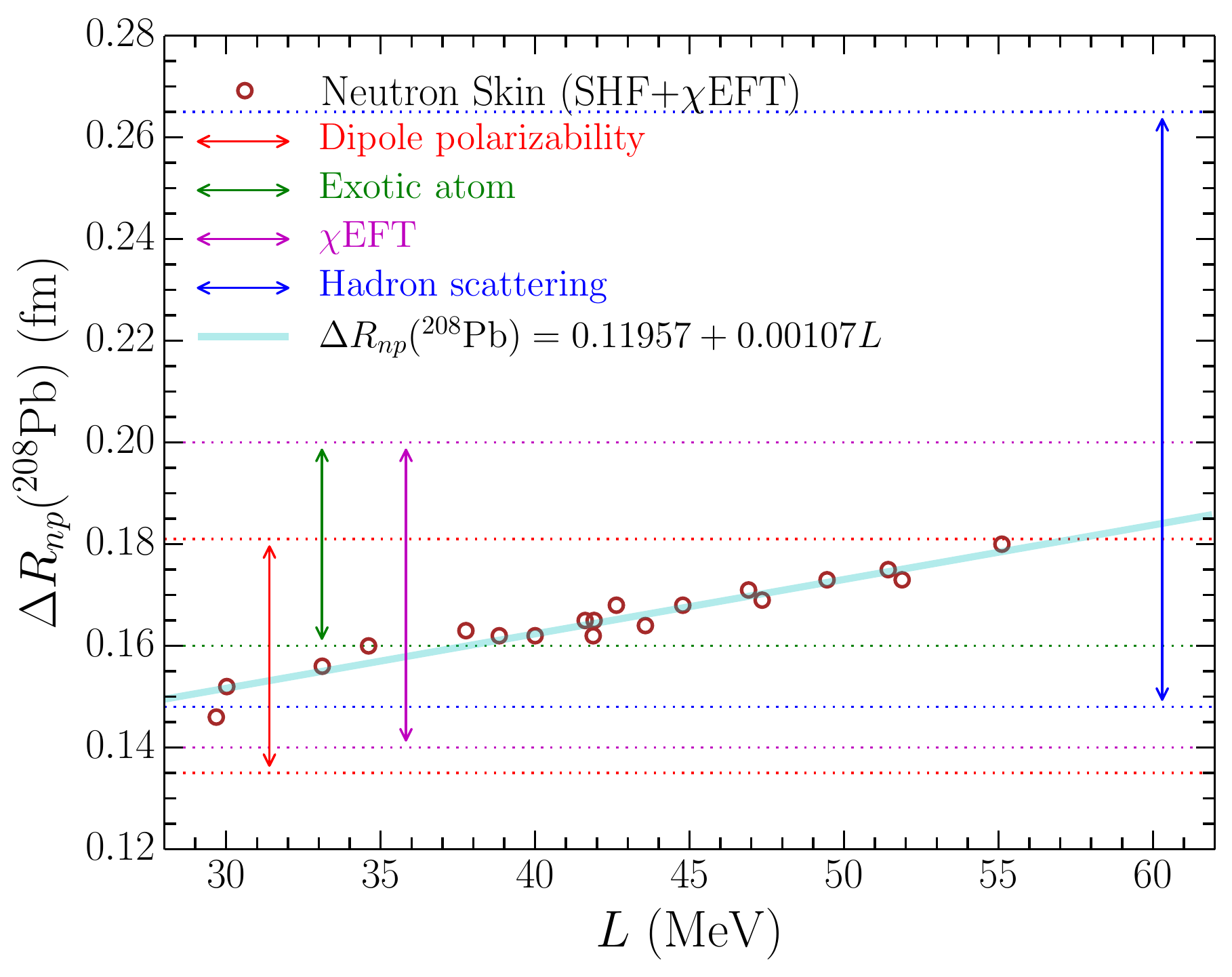}
\caption{Estimates of the neutron skin thickness of $^{208}$Pb from nuclear experiments (described in the text) and mean field models fitted to neutron matter predictions from chiral effective field theory.}
\label{fig:pb208skin}
\end{figure}

Fig.\ \ref{fig:pb208skin} summarizes several of the constraints on the neutron skin thickness of $^{208}$Pb from both nuclear experiment and mean field model calculations. For the mean field theory calculations, we obtained \cite{lim17,zhang18} Skyrme parameters fitted to the neutron matter equation of state from chiral effective field theory \cite{drischler14, holt17prc} and the binding energies of doubly-closed-shell nuclei. From nuclear experiments, we see that the overlapping region of neutron skin has the boundary, $ 0.16 <\Delta R_{np }<0.18\,\mathrm{fm}$, which implies that $40 \le L \le 60\,\mathrm{MeV}$. This range is consistent with the energy density functionals studied in this work, which from the posterior probability distributions have $L = 51 \pm 9$\,MeV. The strong correlation between $L$ and $\Delta R_{np}$ enables us to find a direct correlation between $\Delta R_{np}$ and $\Lambda$ for neutron star deformabilities. 

In the liquid drop model (LDM), the neutron skin thickness is given by \cite{Myers_Swiatecki_1969}
\begin{equation}
\Delta R_{np} = \sqrt{\frac{3}{5}} \left [ t -\frac{e^2}{70}\frac{Z}{J} 
+ \frac{5}{2R}(b_n^2 - b_p^2) \right],
\end{equation}
where $t$ is the distance between the neutron and proton mean surface location, $Z$ is the proton number, $R = r_0 A^{1/3}$ is the nuclear radius, and $b_n$ and $b_p$ are the surface width of neutrons and protons. The quantities $b_n$ and $b_p$ have value $1$\,fm in the standard LDM and $t$ is given by
\begin{equation}
t = \frac{3}{2}r_0 \frac{J}{Q} 
\frac{I -\frac{c_1}{12}\frac{Z}{J}A^{-1/3}}
{1 + \frac{9}{4}\frac{J}{Q}A^{-1/3}},
\end{equation}
where $I = (N-Z)/A$, $Q$ is the surface stiffness coefficient, and $c_1 = \frac{3e^2}{5r_0}$.\\

\begin{figure}[t]
\centering
\includegraphics[scale=0.56]{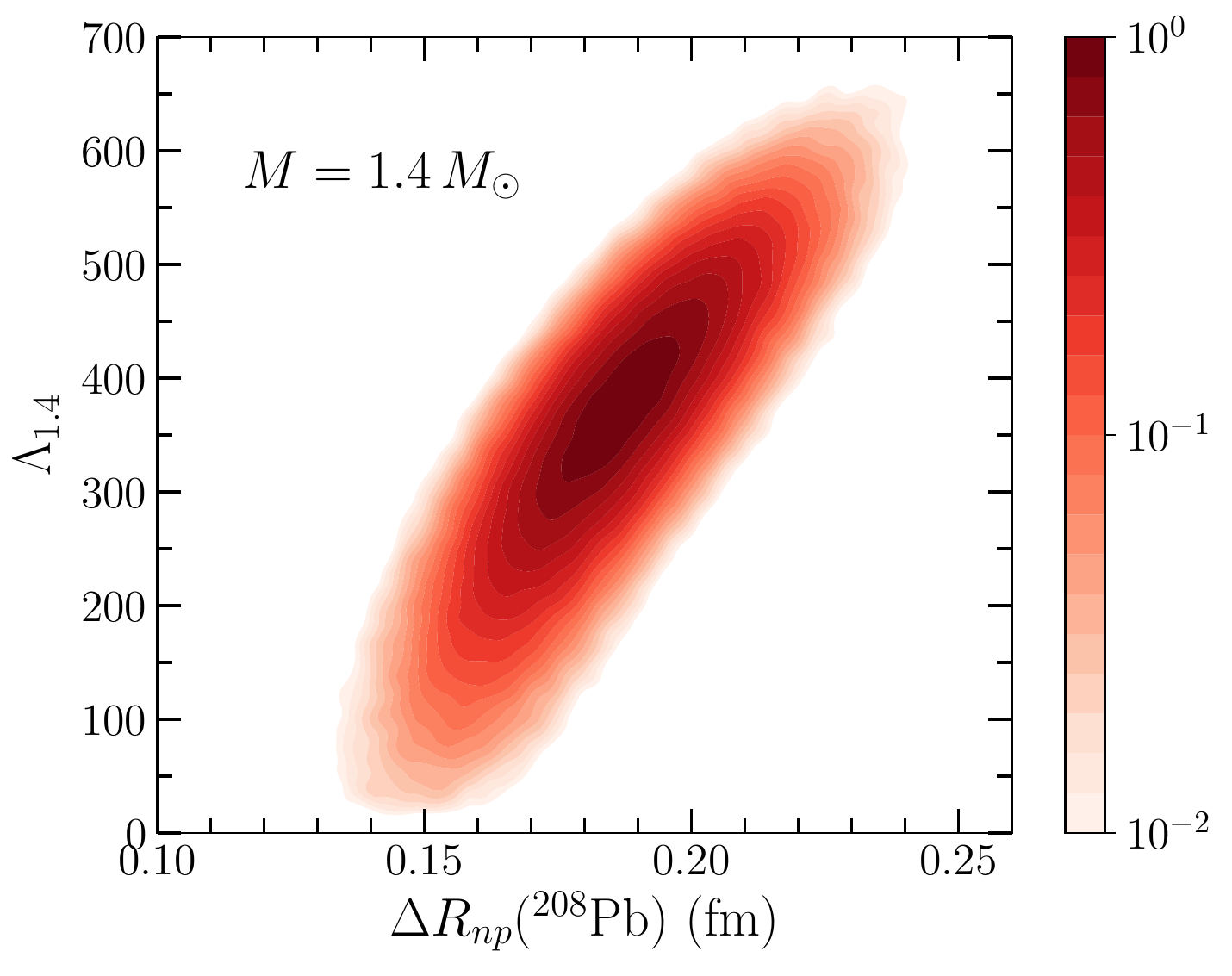}
\caption{Distribution of the $^{208}$Pb neutron skin thickness $\Delta R_{np}$ and tidal deformability $\Lambda_{1.4}$
	of a $1.4\,\msun$ neutron star obtained from the energy density functionals employed in this work.}
\label{fig:tdskin}
\end{figure}

The LDM formula for the neutron skin thickness gives only a rough estimate for nuclei and it deviates from mean field theory calculations by about $10 - 20\%$. A better approximation to the results from mean field theory calculations can be obtained by choosing $J$ and $L$ as independent variables to determine the neutron skin thickness for $^{208}$Pb:
\begin{equation}
\begin{aligned}
\Delta R_{np} ({}^{208}\mathrm{Pb}) = &   (-0.0787 +  0.006736\frac{J}{\mathrm{MeV}}  \\
                & + 0.0009554\frac{L}{\mathrm{MeV} })\,\mathrm{fm}.
\end{aligned}
\end{equation}
This fitting function gives a root-mean-square deviation for the $^{208}$Pb neutron skin thickness $\mathrm{RMSD}(\Delta R_{np})  =9.787\times 10^{-3}$\,fm compared to $\mathrm{RMSD}(\Delta R_{np})$ $= 1.328\times 10^{-2}$\,fm for the fitting function containing only $L$. We used a total of 48 mean field models both from Skyrme Hartree-Fock and relativistic mean field theory covering the range $27 \le J \le 43$\,MeV and $7.17\le L \le 135$\,MeV which is wide enough to represent our energy density functional modeling. Fig.\ \ref{fig:tdskin} shows the two dimensional contour plot of $\Delta R_{np}$ and $\Lambda$. These results also imply that a precise measurement of $\Lambda$ may help to constrain the neutron skin thickness of $^{208}$Pb or vice versa.

In the future we plan to consider a wider range of models for the high-density equation of state, including phase transitions as well
as different powers of the Fermi momentum beyond those in Eq.\ (\ref{eq:exp}). For example, 
we may modify the high-density equation of state by assuming, e.g.,
\begin{subequations}
	\begin{align}
	p & = p_{h_i} \left( \frac{n}{n_{h_i}} \right)^{\Gamma_i} , \\
	\varepsilon & = m_b n + \frac{1}{\Gamma -1 } p_{h_i} \left( \frac{n}{n_{h_i}} \right)^{\Gamma_i}\,
	\end{align}
\end{subequations}
when $n_i\le n \le n_{i+1}$\,. This equation of state can be completed by adding several transition densities $n_i$ and 
corresponding polytropic indices. Since nuclear theory and experiment provide limited insight into the properties of dense nuclear matter beyond $2n_0$, the choice of $n_i$'s and $\Gamma_i$'s can be arbitrary except that such equations of state should reproduce known masses, radii, and tidal deformabilities of neutron stars.


\section{Summary}
\label{sec:sum}

In the present work, we have described a Bayesian approach for implementing constraints on the low- to moderate-density equation of state from nuclear theory and experiment. Microscopic calculations based on high-precision two- and three-body chiral nuclear forces inform our beliefs about the parameters in the equation of state before they are constrained by medium-mass and heavy nuclei experimental data. The latter are incorporated through the Bayesian likelihood function, whose product with the prior probability distribution generates the posterior. The present framework can naturally accommodate future developments in microscopic modeling and rare-isotope experimental data as refinements to the prior and likelihood functions.

Sampling from our Bayesian posterior probability distribution and extrapolating without modification to the high-density regime, we then generate 300,000 equations of state for the statistical analysis of bulk neutron star properties, such as the radius and tidal deformability. The majority of the equations of state are relatively soft, as found in previous works that implemented constraints on the equation of state from chiral effective field theory. At the 95\% credibility level, we find that the radius of a $1.4\,\msun$ neutron star lies in the range $10.0\,{\rm km} < R_{1.4} < 12.7\,{\rm km}$, with the most probable value at $R=12.0$\,km. Similarly, we find that at the 95\% credibility level the tidal deformability of a $1.4\,\msun$ neutron star lies in the range $100 < \Lambda_{1.4} < 500$ with a most probable value of $\Lambda = 350$. These results are consistent with current observational constraints from GW170817.

Future neutron star observations inconsistent with our modeling would require significant modifications to the high-density equation of state. Presently, however, our results are consistent with available astrophysical constraints, except for a fraction ($\sim 30\%$) of equations of state that fail to generate $2.0\,\msun$ neutron stars. As pointed out earlier in the text, this can be remedied by an artificial stiffening of the equation of state beyond twice saturation density $n=2n_0$ while leaving the bulk properties of typical neutron stars with $M \simeq 1.2-1.5\,\msun$ essentially unchanged. The statistical credibility intervals for the pressure, radius, and tidal deformability obtained in the present work can be reduced by future gravitational wave observations from Advanced LIGO and VIRGO, neutron star mass and radius measurements from NICER, a moment of inertia measurement of pulsar J0737-3039A, nuclear experiments involving exotic isotopes, and improved microscopic constraints from chiral effective field theory.

\begin{acknowledgement}
\large\textbf{Acknowledement}\\
\normalsize
\\
We thank Xavier Roca-Maza for discussions. 
Work supported in part by the National Science Foundation under Grant No.\ PHY1652199. 
Portions of this research were conducted with the advanced computing resources provided by 
Texas A\&M High Performance Research Computing.
\end{acknowledgement}

\bibliographystyle{apsrev4-1}
%


\end{document}